\documentclass[journal]{IEEEtran}

\usepackage{amsmath}
\usepackage{graphicx}
\usepackage{subfig} 
\usepackage{tabu}  

\usepackage[colorlinks,linkcolor=blue,anchorcolor=blue,citecolor=green]{hyperref}

\usepackage{dsfont}
\usepackage{stfloats}
\usepackage{amsmath}
\usepackage{cite}
\usepackage{booktabs}
\usepackage{algorithm}
\usepackage{algorithmic}
\usepackage{amssymb}
\ifCLASSINFOpdf

\else
 
\fi

\begin{document}

\title{Change Detection in Multi-temporal VHR Images Based on Deep Siamese Multi-scale Convolutional Neural Networks}

\author{Hongruixuan~Chen,~\IEEEmembership{Student Member,~IEEE,}
        Chen~Wu,~\IEEEmembership{Member,~IEEE,}
        Bo~Du,~\IEEEmembership{Senior~Member,~IEEE,}
        and~Liangpei~Zhang,~\IEEEmembership{Fellow,~IEEE}% <-this % stops a space

\thanks{Manuscript submitted July 8, 2020.}
\thanks{H. Chen is with the State Key Laboratory of Information Engineering in Surveying, Mapping and Remote Sensing, Wuhan University, Wuhan, P.R. China (e-mail:Qschrx@whu.edu.cn).}
\thanks{C. Wu is with the State Key Laboratory of Information Engineering in Surveying, Mapping and Remote Sensing, Wuhan University, Wuhan, P.R. China (e-mail: chen.wu@whu.edu.cn, corresponding author).}% <-this % stops a space
\thanks{B. Du is with the School of Computer Science, and Collaborative Innovation Center of Geospatial Technology, Wuhan University, Wuhan, P.R. China (email: gunspace@163.com).}% <-this % stops a space
\thanks{L. Zhang is with the Remote Sensing Group, State Key Laboratory of Information Engineering in Surveying, Mapping, and Remote Sensing, Wuhan University, Wuhan, P.R. China (e-mail: zlp62@whu.edu.cn).}
}

\markboth{SUBMITTED TO XXXX}%
{Shell \MakeLowercase{\textit{et al.}}: Bare Demo of IEEEtran.cls for IEEE Journals}

\maketitle

\begin{abstract}
  Very-high-resolution (VHR) images can provide abundant ground details and spatial geometric information. Change detection in multi-temporal VHR images plays a significant role in urban expansion and area internal change analysis. Nevertheless, traditional change detection methods can neither take full advantage of spatial context information nor cope with the complex internal heterogeneity of VHR images. In this paper, a powerful feature extraction model entitled multi-scale feature convolution unit (MFCU) is adopted for change detection in multi-temporal VHR images. MFCU can extract multi-scale spatial-spectral features in the same layer. Based on the unit two novel deep siamese convolutional neural networks, called as deep siamese multi-scale convolutional network (DSMS-CN) and deep siamese multi-scale fully convolutional network (DSMS-FCN), are designed for unsupervised and supervised change detection, respectively. For unsupervised change detection, an automatic pre-classification is implemented to obtain reliable training samples, then DSMS-CN fits the statistical distribution of changed and unchanged areas from selected training samples through MFCU modules and deep siamese architecture. For supervised change detection, the end-to-end deep fully convolutional network DSMS-FCN is trained in any size of multi-temporal VHR images, and directly outputs the binary change map. In addition, for the purpose of solving the inaccurate localization problem, the fully connected conditional random field (FC-CRF) is combined with DSMS-FCN to refine the results. The experimental results with challenging data sets confirm that the two proposed architectures perform better than the state-of-the-art methods.
\end{abstract}

\begin{IEEEkeywords}
  Change detection, very-high-resolution images (VHR images), multi-temporal images, multi-scale feature convolution, deep siamese convolutional neural network, fully connected conditional random field (FC-CRF)
\end{IEEEkeywords}

\IEEEpeerreviewmaketitle

\section{Introduction}\label{sec:1}

\IEEEPARstart{C}{hange} detection is one of the major and hot topics in the remote sensing field. The most commonly used definition of change detection is given by Singh \cite{Singh1989}: change detection is the process of identifying differences in the state of an object or phenomenon by observing it at different times. Now, change detection is widely applied to ecosystem monitoring, resource management, land-use/land-cover change analysis, urban expansion research, and damage assessment \cite{Xian2009, Coppin2004, Luo2018, Lu2004, Brunner2010, Zelinski2014}.

\par Multi-spectral images with medium- and low- spatial resolution are the most commonly used data source for change detection, thus numerous change detection methods have been developed and newer methods are still emerging. Change vector analysis (CVA) is one of the most classic change detection methods that generates a difference image (DI) and performs clustering algorithms in DI to obtain the change detection result \cite{Sharma2007}. CVA is also the backbone for some of the more advanced change detection methods \cite{Bovolo2012, Thonfeld2016, Saha2019}. As a dimension reduction method, principal component analysis (PCA) uses an orthogonal transformation to convert original images into a new orthogonal feature space and chooses a part of principal components for change detection \cite{Deng2008}. Multivariate alteration detection (MAD) and its iterative version iteratively re-weighted MAD (IRMAD) extract changed pixels by maximizing the difference between the transformed variables \cite{Nielsen1997, Nielsen2007}. A novel change detection method is proposed by Wu et al. in \cite{Wu2014, Wu2017a}. This method based on slow feature analysis (SFA) theory tries to seek the most invariant components in multi-temporal images for change detection.

\par Nowadays, with the development of Earth observation technology, VHR images are more available by lots of satellite sensors, such as IKONOS, Worldview, SPOT, GaoFen (GF), and QuickBird. VHR images have abundant detailed geometric information, which has crucial effects on the research of urban change analysis and building detection \cite{Tang2011,Wen2016,Wu2017a,Luo2018}. Therefore, VHR images change detection has caught more and more attention in the remote sensing field \cite{Lei2014,wu2019unsupervised,chen2020dsdanet,Chen2019a,Tan2016, Hussain2013, Chen2012,Somasundaram2010,Benedek2009,Moser2011,Hoberg2015,Zhou2016,Lv2018a,Zhan2017,Saha2018,Saha2020a,Saha2020b}. 

\par Nevertheless, as spatial resolution increases, the internal heterogeneity of the same class also increases. The aforementioned change detection methods for low- and medium- spatial resolution images only explore spectral features and neglect spatial context information, thus they may not apply for change detection in VHR images. To achieve better change detection performance, a lot of methods are designed to utilize spatial-spectral features. In \cite{Lei2014}, a method based on texton forest is developed to capture spatial context information. Tan et al explore texture and morphological profiles in VHR images for change detection \cite{Tan2016}. Based on the assumption that the spectral vectors of the pixels belonging to the same type object are similar, some object-based change detection (OBCD) methods are designed and achieve relatively good results \cite{Hussain2013,Chen2012,Somasundaram2010}. Besides, some probabilistic graph models, such as conditional random field (CRF) \cite{Sutton2012} and Markov random field (MRF) \cite{Li1994}, are introduced to utilize spatial context information for change detection \cite{Benedek2009,Moser2011,Hoberg2015,Zhou2016,Lv2018a}. However, these methods only extract low-level features from VHR images for change detection, which are not robust and insufficient for representing the key information of original images.

\par Recently, deep learning (DL) has achieved significant performance in many domains \cite{Lecun2015}, including remote sensing image interpretation \cite{Zhang2016b,Zhu2017a}. Convolutional neural network (CNN), as a classical and powerful DL architecture has the capacity to capture multi-level features in an automatic manner \cite{Y1998}, which is suitable for processing VHR images. Consequently, a variety of methods based on CNN have been proposed for change detection in multi-temporal VHR images. Saha et al. \cite{Saha2019} propose an unsupervised method called deep change vector analysis (DCVA) for binary and multi-class change detection. In DCVA, a pre-trained deep CNN is adopted to extract deep spatial-spectral features from multi-temporal images. In \cite{Liu2018}, a deep convolutional neural network entitled symmetric convolutional coupling network (SCCN) is introduced for change detection in heterogeneous images. In SCCN, the convolutional layer is responsible for feature extraction. For multi-temporal aerial image change detection, Zhan et al. \cite{Zhan2017}  design a deep siamese convolutional network, which extracts features through two weight-shared branches. In \cite{CayeDaudt2018}, the fully convolutional siamese network is first introduced into change detection and three networks are designed. The three networks are trained in an end-to-end manner and achieve good performance. For the purpose of extracting high-level spatial-spectral features from multi-source VHR images, a deep siamese convolutional neural network is proposed in \cite{Chen2019}. After high-level spatial-spectral features are extracted, a multiple-layers recurrent neural network is designed to mine change information. Except for CNN architecture, a change detection method based on conditional generative adversarial network (cGAN) is designed and a convolutional layer aims to extract spatial-spectral features from VHR images \cite{Niu2019}. In this method, the convolutional layer is responsible for extracting features from VHR image patches. 

\par All of these above methods adopt a single scale convolution kernel as the feature extraction module. Although the single scale convolution kernel could extract spatial-spectral features to a certain degree, it still has some powerlessness in coping with the complex ground situations of VHR images. Few research has attempted to explore other sizes of convolution kernels or even multiple convolution kernels for change detection in VHR image. What’s more, unsupervised and supervised change detection often face different scenarios, it is hard to find a general architecture that is suitable for both tasks simultaneously. On the one hand, in supervised change detection, these network structures proposed for unsupervised change detection, such as SCCN, have limited fitting ability, which makes it difficult to obtain accurate change detection result. On the other hand, the architectures developed for supervised change detection cannot be adequately trained in unsupervised tasks, also resulting in unsatisfactory results. Therefore, it is necessary to design different network architectures for unsupervised and supervised change detection, respectively.

\par Considering the above issues comprehensively and inspired by Inception network \cite{Szegedy2015a}, the multi-scale feature convolution unit proposed in \cite{Szegedy2015a} is adopted to extract multi-scale spatial-spectral features in the same layer, which is suitable for processing VHR images. Adopting MFCU as the basic feature extraction module, two powerful deep siamese convolutional neural networks are designed for unsupervised and supervised change detection, respectively. For supervised change detection, a probabilistic graph model, FC-CRF \cite{Krahenbuhl2011a} is adopted to refine the change detection results. Based on the two networks and FC-CRF, the specific supervised and unsupervised algorithms are developed.

\par The contributions of this paper \footnote{This paper is an improved version of the original conference paper: https://ieeexplore.ieee.org/document/8866947} are summarized as follows:

\begin{enumerate} 
  \item This paper introduces a multi-scale feature convolution unit into change detection, which has the capacity to extract multi-scale spatial-spectral features from VHR images. What’s more, the unit is a flexible module and can be used in any deep neural networks designed for the tasks involving VHR images. To the authors’ best knowledge, this is the first time that such multi-scale feature convolution unit is exploited for change detection.
  \item For the fact that unsupervised and supervised change detection often face different scenarios, based on MFCU, two novel deep siamese convolutional neural networks are developed for unsupervised and supervised change detection, respectively. Among them, the network used for supervised change detection is able to process images of any size and does not require sliding patch-window, thus the accuracy and inference speed could be significantly improved.
  \item In the proposed supervised change detection algorithm, for the purpose of solving the problem of inaccurate localization caused by deep convolution architecture, FC-CRF is adopted to refine the results obtained by DSMS-FCN.
\end{enumerate}

The rest of this paper is organized as follows. Section \ref{sec:2} introduces the MFCU, DSMS-CN, and specific unsupervised change detection algorithm in detail. In section \ref{sec:3}, the DSMS-FCN, FC-CRF, and corresponding supervised change detection algorithm are described. To evaluate the proposed methods, the experiments of unsupervised and supervised change detection are carried in section \ref{sec:4} and section \ref{sec:5}, respectively. In the end, Section \ref{sec:6} draws the conclusion of our work in this paper.

\begin{figure}[t]

  \centering
  \includegraphics[scale=0.5]{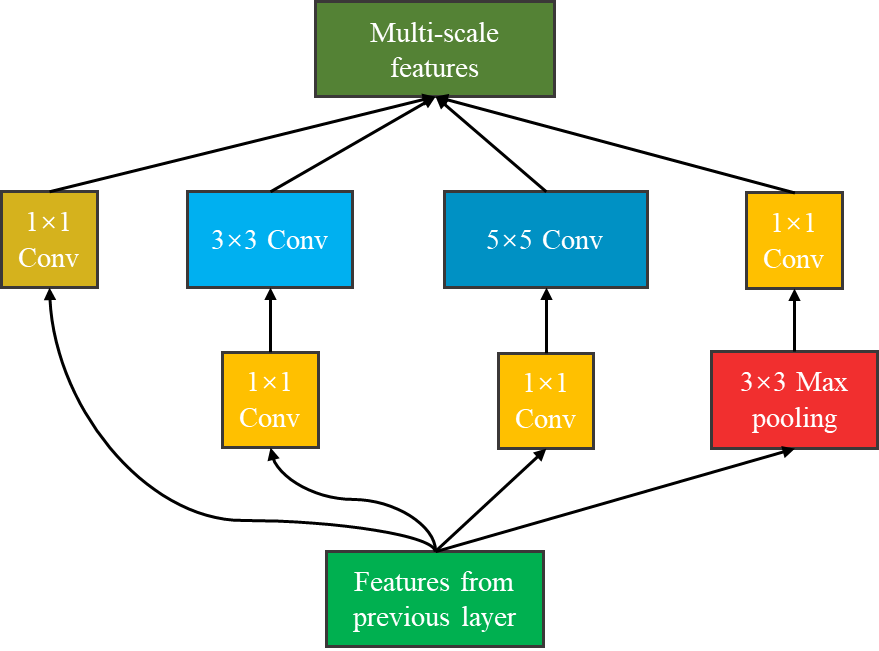}
  \caption{Illustration of MFCU. Different from the conventional single-scale convolution unit, MFCU can extract multi-scale spatial-spectral features in parallel by four ways in the same layer.}
  \label{fig_MFCU}
\end{figure}
\section{Unsupervised Change Detection}\label{sec:2}
In this section, MFCU, the network designed for unsupervised change detection and specific change detection algorithm are elaborated. Though MFCU is introduced in this section, it is also the basic module of the network proposed for supervised change detection.
\subsection{Multi-scale Feature Convolution Unit}
\par The VHR images can provide abundant ground details, texture information, and spatial distribution information \cite{Chen2019}. As a commonly used convolution unit, 3$\times$3 convolution kernel could extract spatial-spectral features from VHR images. However, the 3$\times$3 convolution kernel has two obvious disadvantages in feature extraction of VHR images. First, the 3$\times$3 convolution kernel could only extract single scale spatial-spectral features in the same layer. But in VHR images, there exist different features, varying from small scales to large scales. Besides, as a weighted summation operation, convolution has a smoothing effect, thus some changes existing in multi-temporal images could be erased. Therefore, it is undeniable that the 3$\times$3 convolution kernel, or say, the conventional single-scale convolution unit is somewhat incompetent in dealing with complex multi-scale ground conditions in VHR images. 

\par Therefore, to extract multi-scale spatial-spectral features, the Inception module first proposed in \cite{Szegedy2015a} is adopted, and we rename it as MFCU in this paper. As illustrated in Fig. \ref{fig_MFCU}, MFCU has a “network in network” structure \cite{Lin2013} and can extract multi-scale spatial-spectral features by 1$\times$1 convolution kernel, 3$\times$3 convolution kernel, 5$\times$5 convolution kernel, and 3$\times$3 max pooling, respectively. In the above four ways, the 1$\times$1 convolution kernel concentrates on extracting the features of a pixel itself. The 3$\times$3 convolution kernel is able to extract the spatial-spectral features. The 5$\times$5 convolution kernel extracts spatial-spectral features over a larger range, which applies for a few large-scale continuous objects in VHR images. And the 3$\times$3 max pooling could extract the most salient features and efficiently avoids the smoothing effect of the convolution operation. At last, the four type features are fused to obtain the higher dimensional multi-scale features. In addition, a bottleneck design \cite{Simonyan2014} is adopted in MFCU, which uses the 1$\times$1 convolution kernel to reduce the feature dimensions. This structure can efficiently reduce the number of parameters and make the training process of network easier.

\par Compared with the conventional single-scale convolution unit, MFCU could extract multi-scale features, which makes the feature extraction ability of network more powerful and does not increase parameters of network.

%\IEEEpubidadjcol

\subsection{Deep Siamese Multi-scale Convolutional Network}

\begin{figure}[t]

  \centering
  \includegraphics[scale=0.65]{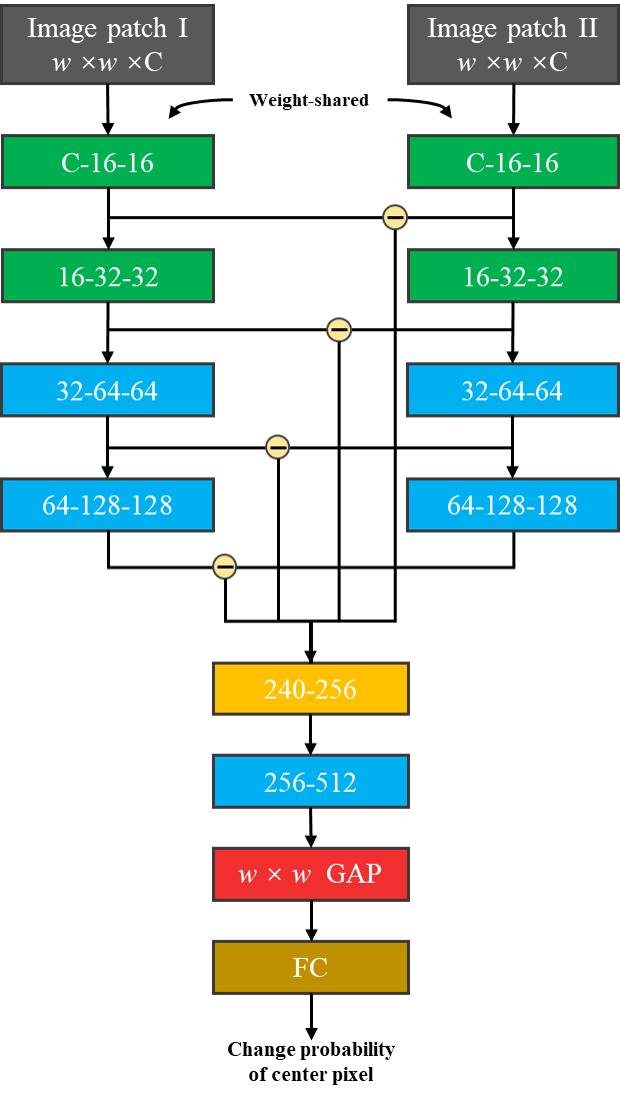}
  \caption{Architecture of the proposed DSMS-CN. DSMS-CN extracts features from multi-temporal VHR image patches with a fixed size and outputs change probability of center pixels. Legend: green block indicates a conventional 3$\times$3 convolution module, blue block means the MFCU module, yellow block is an 1$\times$1 convolution module, red block is a w$\times$w global average pooling layer and brown block means a fully connected layer. The numbers in each block represent the change of the number of feature channels in each module.}
  \label{fig_DSMSCN}
\end{figure}

\par Using MFCU as a basic feature extraction module, the network entitled DSMS-CN is illustrated in Fig. \ref{fig_DSMSCN}, which consists of two sub-networks: feature extraction network and change judging network. DSMS-CN processes image patches with a fixed size and outputs change probabilities of the center pixels.

\par The feature extraction network of DSMS-CN is a siamese convolutional network \cite{Bromley1993}. Its two branches extract spatial-spectral features from two multi-temporal image patches with a fixed size using exactly the same way. In each branch, the former two conventional convolutional modules transform the original image patches into relative high dimensional representation, then abundant multi-scale features are extracted by the two latter MFCU modules. To highlight change information, the absolute values of feature differences are calculated \cite{Saha2019}. Then the feature differences with different level are concatenated and a 1$\times$1 convolution layer is used to fuse these features. 

\par In the change judging network, multi-scale features are further extracted by an MFCU first. For the purpose of making the network more robust and mitigating overfitting \cite{Lin2013}, a global average pooling layer (GAP) replaces the fully connected layer to generate feature vector. Finally, the change probability of the center pixel in each patch is obtained by a fully connected layer. 

\par In DSMS-CN, the activation functions of both conventional and multi-scale convolutional layers are rectified linear units (ReLU), the sigmoid function is adopted in the last fully connected layer to predict the change probability. For the purpose of preserving the information to the largest degree, DSMS-CN does not adopt the max-pooling layer to reduce dimension. 

\begin{figure*}[t]

  \centering
  \includegraphics[scale=0.75]{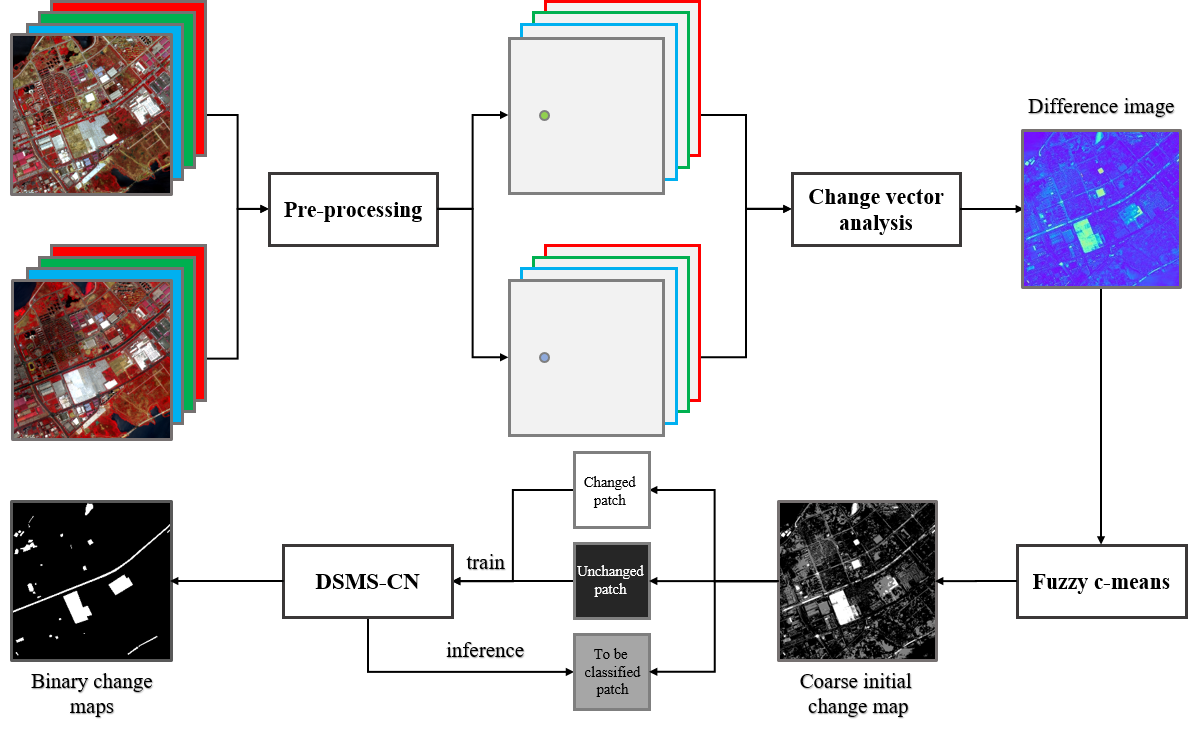}
  \caption{Flowchart of the unsupervised change detection alogrithm. The co-registration and radiometric correction are first implemented to make the geometry and radiometric conditions of multi-temporal VHR images consistent. Then suitable training image patches are chosen by CVA and FCM. Finally, training DSMS-CN with the selected image patches, and then DSMS-CN classifies the center pixels of the remaining image patches to obtain full change detection result. Note that the change detection decision made by DSMS-CN is at the pixel level instead of the patch level.}
  \label{fig_unsupervised}
\end{figure*}

\subsection{Unsupervised Change Detection Algorithm}

\par Adopting DSMS-CN to obtain change detection results, the specific unsupervised change detection algorithm is presented. The pipeline of the algorithm is illustrated in Fig. \ref{fig_unsupervised}. The first step is image pre-processing, including co-registration and radiometric correction. Image registration is defined as “the process of aligning two or more images of the same scene acquired at diverse times” \cite{Zitova2003}. Through collecting matched point-pairs, constructing transform models and transforming images, the multi-temporal VHR images are geometrically aligned. Then radiometric relative normalization is implemented to reduce the radiometric difference between multi-temporal VHR images caused by distinct imaging conditions, such as light intensity, sun zenith angle, and atmospheric conditions. The specific implementation is normalizing the multi-temporal VHR images with zero mean and unit variance, respectively.

\par After pre-processing images, the suitable training samples are chosen by automatic pre-classification. The main purpose of this step is to choose reliable samples for training the proposed network. CVA is first adopted to calculate DI of multi-temporal VHR images. Then, fuzzy c-means clustering algorithm (FCM), based on the memberships of pixels, is performed to partition DI into three clusters: $w_{c}$, $w_{uc}$ and $w_{tbc}$. The pixels belonging to $w_{c}$ and $w_{uc}$ are reliable pixels that have high change and non-change probabilities, respectively. And the pixels in $w_{tbc}$ are unreliable and need to be classified. The $w \times w$ neighborhood area of pixels in $w_{c}$ and $w_{uc}$ are chosen as training samples. This automatic sample selection method based on pre-classification has been utilized in many change detection models \cite{Gao2016, Gong2017a, Gong2017b, Li2019, Geng2019, Li2019a, Du2019a, Liu2019b, Song2020a}. Besides, some advanced change detection models, such as DCVA \cite{Saha2019}, based on CVA are developed.

\par Eventually, DSMS-CN is trained on the chosen training image patches. After the training process is completed, the pixels in $w_{tbc}$ are detected by DSMS-CN and the full change detection result is generated. Owing to sigmoid function of fully connected layer of DSMS-CN, the threshold segmentation step could be simplified and just sets threshold as 0.5 to get binary change map. Though training the proposed DSMS-CN is a supervised learning process, the entire process of the algorithm is an unsupervised fashion without any priori knowledge.

\begin{figure*}[ht]

  \centering
  \includegraphics[scale=0.6]{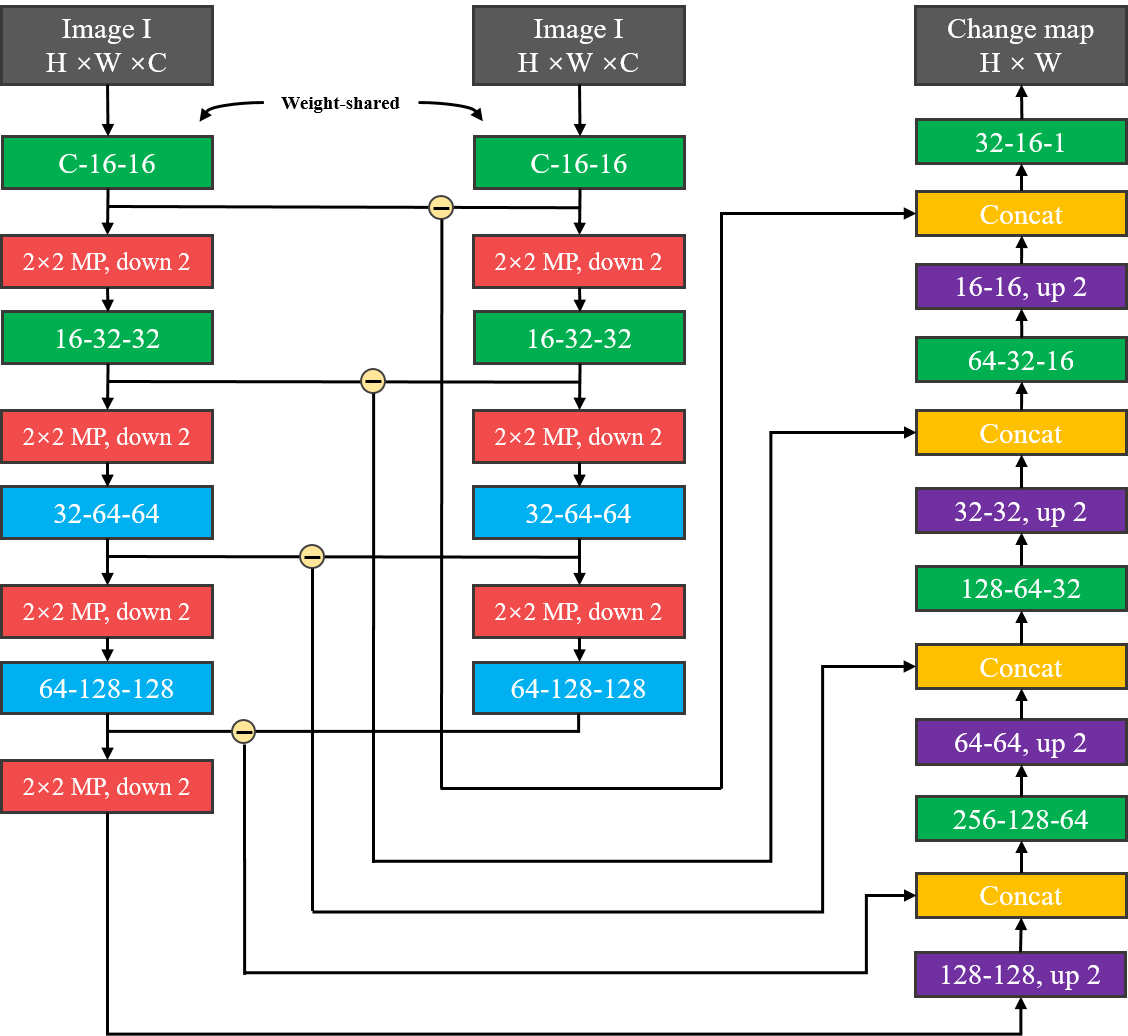}
  \caption{Architecture of the proposed DSMS-FCN. The encoder network directly processes multi-temporal VHR images to get high dimensional multi-scale feature maps and the decoder network generates change maps based on feature difference from multiple layers and high dimensional multi-scale feature map of one branch. Legend: green block indicates a conventional 3$\times$3 convolution module, blue block means the MFCU module, red block module is a 2$\times$2 max-pooling layer and purple block module is a transpose convolution module. The numbers in each block represents the change of the number of feature channels.}
  \label{fig_DSMSFCN}
\end{figure*}

\section{Supervised Change Detection}\label{sec:3}
In this section, the network proposed for supervised change detection, FC-CRF, and specific supervised change detection algorithm are introduced.
\subsection{Deep Siamese Multi-scale Fully Convolutional Network}
\par Based on MFCU (for details, please refer to Section II-A), the proposed network for supervised change detection is a fully convolutional network entitled DSMS-FCN. DSMS-FCN consists of an encoder network and a decoder network. The specific architecture of DSMS-FCN is shown in Fig. \ref{fig_DSMSFCN}.

\par The encoder network of DSMS-FCN has two weight-shared branches. Each branch has four subsampling layers and each subsampling layer includes a convolution module and a 2$\times$2 max-pooling layer. The former two convolution modules consist of 3$\times$3 convolution kernels and the latter two modules consist of MFCUs. Based on the skip-connection structure proposed in the U-Net \cite{Ronneberger2015}, the features in subsampling layer and upsampling layer at the same scale are concatenated during the upsampling phase, which could recover localization information to a certain degree and generate more accurate binary change maps with precise boundaries. The motivation for concatenating the absolute values of feature differences with the features in the upsampling layer is that change detection aims at detecting the differences between multi-temporal images. Furthermore, it is worth pointing out that only one branch’s features enter into decoder network. This is because the two branches share weights and changes in multi-temporal VHR images are only in the minority, thus most of features extracted from both sides are same. Meanwhile, feature differences are delivered to decoder network through the skip-connection structure. Consequently, using features of only one branch could avoid feature redundancy and require a smaller number of parameters.

\par In DSMS-FCN, all convolutional layers and transpose convolutional layers adopt ReLU as activation function, except the last convolutional layer, which adopts sigmoid function to predict change probability of the whole image. Owing to the fully convolutional architecture of DSMS-FCN, it can process multi-temporal VHR images of any size.

\begin{figure*}[t]

  \centering
  \includegraphics[scale=0.55]{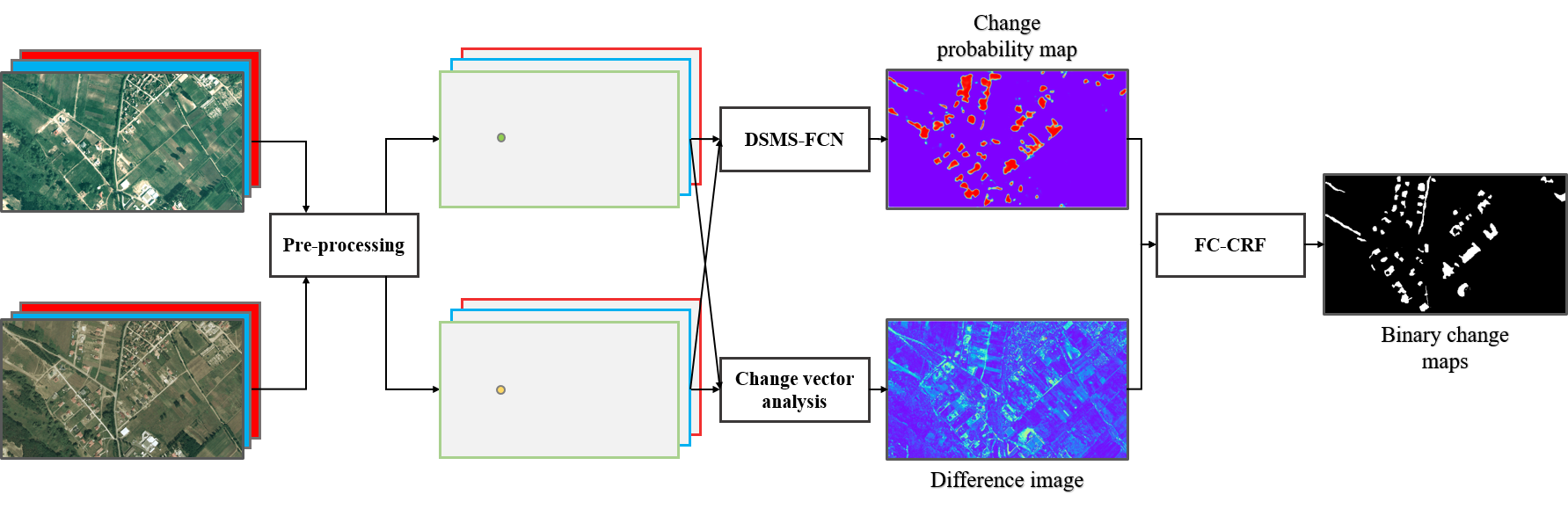}
  \caption{Flowchart of the supervised change detection alogrithm. The pre-processing is first performed on the data set. DSMS-FCN and FC-CRF are trained on the training set. After the training step is completed, DSMS-FCN infers change probability map of given multi-temporal VHR images. Then the FC-CRF refine the results obtained by DSMS-FCN depending on the change probability map and DI acquired by CVA. At last, the FC-CRF generates a more precise binary change map.}
  \label{fig_supervised}
\end{figure*}

\subsection{Fully Connected Conditional Random Field}
\par Though DSMS-FCN adopts the skip-connection structure to deliver localization information, it still suffers from the problem of inaccurate localization caused by invariance of features and large receptive field \cite{Chen2018}. To tackle this problem, FC-CRF \cite{Krahenbuhl2011a} is adopted to refine the localization information of the results obtained by DSMS-FCN. Compared with CRF, FC-CRF considers short-range and long-range information simultaneously, thus it can better recover the local structure. 

\par FC-CRF is a conditional probability distribution model that outputs another set of random variables given a set of input random variables. According to \cite{Krahenbuhl2011a}, the energy function of FC-CRF is defined as follows:

\begin{equation}
  E(Y|X) = \sum\limits_{i}\phi_{u}(y_{i}) + \sum\limits_{i<j}\phi_{p}(y_{i},y_{j})
\end{equation} 
where $i$ and $j$ range from 1 to $N$, $\phi_{u}$ indicates unary potential, and $\phi_{p}$ indicates pair-wise potential. In change detection, $X=\{ x_{1},x_{2},\cdots,x_{N}\}$ is the observed image acquired by the difference of multi-temporal images 
and $Y=\{ y_{1},y_{2},\cdots,y_{N}\}$ is the binary change map.

\par The domain of each $y_{i}$ is $L= \{ 0,1 \}$. The unary potential 

\begin{equation}
  \phi_{u}(y_{i})=-logP(y_{i})
\end{equation} 

\par where $P(y_{i})$ is change probability of pixel $i$, which is computed by DSMS-FCN. In addition, each pixel pair has a corresponding pairwise term no matter how far apart they are. The pairwise $\phi_{p}(y_{i},y_{j})$ potential has the form:
\begin{equation}  
  \left\{\begin{matrix}
    \phi_{p}(y_{i},y_{j})=\mu(y_{i},y_{j})k(f_{i},f_{j}) \\
    k(f_{i},f_{j})=\sum\limits_{m=1}^{n}w^{(m)}k^{(m)}(f_{i},f_{j}) \\ 
    \end{matrix}\right.    
\end{equation}  
\par Here, $\mu$ is a penalty, $\mu(y_{i},y_{j})=1 \ if \ y_{i}\not=y_{j}$ and zero otherwise.$k^{(m)}$ is a Gaussian kernel and is weighted by $w^{(m)}$ and $n$ is the number of kernels. $f_{i}$ and $f_{j}$ are feature vectors for pixels $i$ and $j$ in a feature space.

\par In change detection problem, the kernels are

\begin{equation}  
  \begin{split}
  k(f_{i},f_{j})=
  w_{1}exp(-\frac{\parallel c_{i}-c_{j} \parallel_{2}^{2}}{2\sigma_{\alpha}^{2}}
  -\frac{\parallel d_{i}-d_{j} \parallel_{2}^{2}}{2\sigma_{\beta}^{2}})\\
  + w_{2}exp(-\frac{\parallel c_{i}-c_{j} \parallel_{2}^{2}}{2\sigma_{\gamma}^{2}})
  \end{split}
\end{equation}
\par where the first kernel depends on both pixel co-ordinates (denoted as $c$) and spectral difference intensities (denoted as $d$). The second smoothness kernel only depends on pixel co-ordinates. 

\par The inference of FC-CRF adopts mean field approximation algorithm (MFA), , which can be significantly accelerated by high dimensional filter algorithm \cite{Krahenbuhl2011a}.

\subsection{Supervised Change Detection Algorithms}

\par Based on the proposed DSMS-FCN and FC-CRF, the pipeline of the supervised change detection algorithm is shown in Fig. \ref{fig_supervised}. Same as unsupervised architecture, the first step is pre-processing, which can eliminate the difference of geometry and radiometric conditions of multi-temporal VHR images. Then, DSMS-FCN is trained on change detection data sets in an end-to-end manner. The inputs of the DSMS-FCN are a pair of multi-temporal VHR images, and the output is a corresponding change probability map. Different from the proposed DSMS-CN and majority of patch-based models \cite{Zhan2017, Liu2018, Chen2019}, DSMS-FCN can process images of any size and does not require sliding patch-window. Therefore, the change detection accuracy and inference speed could get significantly improved \cite{Shelhamer2017}. 

\par As mentioned before, skip-connection structure of DSMS-FCN cannot solve the problem of inaccurate localization, thus we make use of FC-CRF to refine the results obtained by DSMS-FCN. The parameters of FC-CRF is trained on training set or validation set and the training process between FC-CRF and DSMS-FCN are decoupled. Firstly, the DI is computed with CVA and the change probability map of multi-temporal VHR images is inferred by DSMS-FCN. Then the unary potential of FC-CRF is generated by change probability map and the pairwise potential of FC-CRF is computed by DI. Based on its fully connected structure, the FC-CRF can efficiently extract accurate localization information by considering short-range and long-range information simultaneously. Consequently, the FC-CRF can further refine the results obtained by DSMS-FCN and eventually obtain a better binary change map with more accurate boundaries. Because of adopting high dimension filter algorithm to accelerate MFA, the inference of FC-CRF is very fast in practice. Therefore, the entire algorithm may take some time during training phase, but the speed of inference is still fast.

\begin{figure*}[ht]
  \centering

  \subfloat[]{
    \includegraphics[width=2.2in]{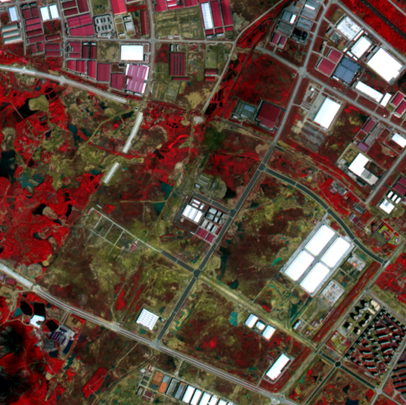}
  \label{WH_dataset_a}}
  \hfil
  \subfloat[]{
    \includegraphics[width=2.2in]{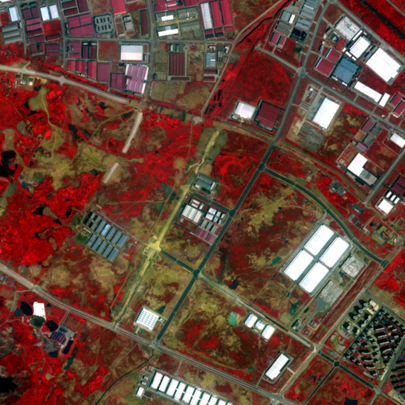}
  \label{WH_dataset_b}}
  \hfil
  \subfloat[]{
    \includegraphics[width=2.2in]{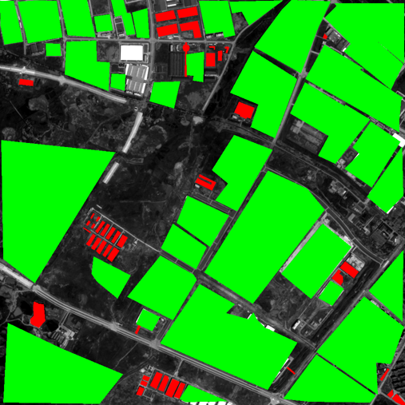}
  \label{WH_dataset_GT}}

  \caption{WH data set. (a) Pre-change. (b) Post-change. (c) is ground truth where red means change and green indicates non-change.}
  \label{WH_dataset}

\end{figure*}

\begin{figure*}[ht]
  \centering

  \subfloat[]{
    \includegraphics[width=2.2in]{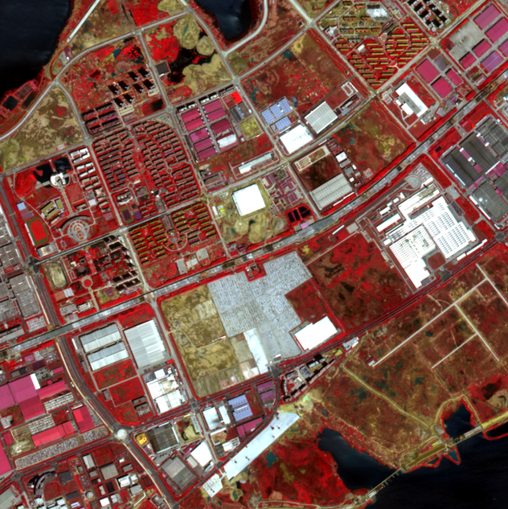}
  \label{fig_first_case}}
  \hfil
  \subfloat[]{
    \includegraphics[width=2.2in]{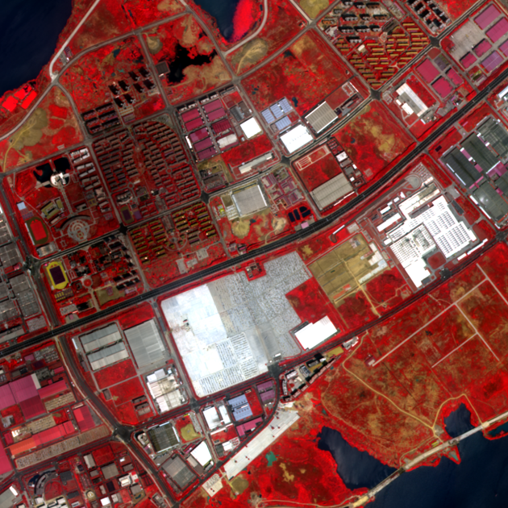}
  \label{fig_second_case}}
  \hfil
  \subfloat[]{
    \includegraphics[width=2.2in]{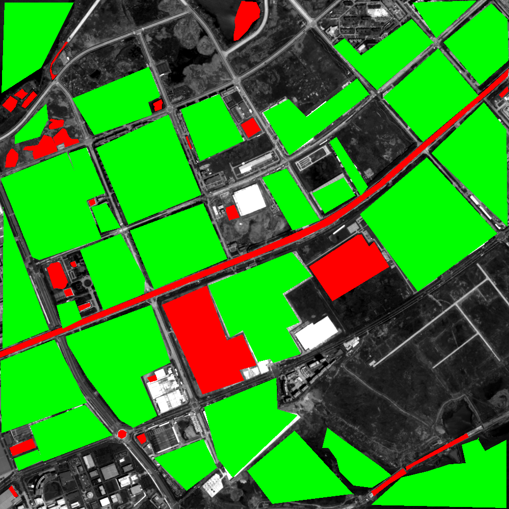}
  \label{fig_third_case}}

  \caption{HY data set. (a) Pre-change. (b) Post-change. (c) is ground truth where red means change and green indicates non-change.}
  \label{HY_dataset}

\end{figure*}

\section{Unsupervised Change Detection Experiment}\label{sec:4}
\subsection{Data Set}

\par In the unsupervised change detection experiment, the first VHR data set called as WH was captured by GaoFen-2 (GF-2) sensor on April 4, 2016 and September 1, 2016, covering the city of Wuhan, China. The image size is 1000$\times$1000 with four bands consisting of red, green, blue and near-infrared. Its spatial resolution is 4 m. Fig. \ref{WH_dataset} shows the pseudo-color images and ground truth of change and non-change. (a) and (b) are the pseudo-color images acquired on April 4, 2016 and September 1, 2016, respectively. (c) is the ground truth. The changed area (red) contains 20026 pixels, and the unchanged area (green) contains 484143 pixels. The remaining pixels are undefined. 

\par The second data set is HY data set with image size of 1000$\times$1000, the two multi-temporal VHR images in this data set were also acquired by GF-2. The images cover the Hanyang city. Fig. \ref{HY_dataset} shows the pseudo-color images and ground truth. The changed area (red) contains 59051 pixels, and the unchanged area (green) contains 416404 pixels. 

\par It could be observed that in both data sets, the changed area only occupies a small part, thus there exists a heavy skewed-class problem between changed and non-change classes, which brings greater challenge to change detection. In addition, there exists the “over-exposed” problems on some buildings in VHR images, which break the linear relationship of radiometric intensity of unchanged regions between multi-temporal images and cannot be eliminated by radiometric normalization \cite{DHINGRA2015738,Luo2018}. Hence, the “over-exposed” problem makes accurate change detection more difficult.

\subsection{Experiment Settings}
\par Firstly, the weights and bias of DSMS-CN are initialized by “he-normal” way \cite{He2015Delving}. In order to overcome the skew-class problem, weighted binary cross-entropy (WBCE) function is applied as the loss function of DSMS-CN:
\begin{equation}  
  L=w_{p}\hat{y}logy+(1-\hat{y})log(1-y),
\end{equation}
where $w_{p}$ is the reciprocal of the proportion of non-change and change classes in training samples. Through setting a larger weight for change class, the change samples would play a more important role at training phase. Adam optimizer \cite{Kingma2014} is chosen to train the network (learning rate is set to 1e-4). Dropout \cite{Nitish2014} and weight decay \cite{Moody1995} are used to avoid overfitting during training phase. The image patch size for DSMS-CN is set as 13 for both data sets. The specific influence at different values is discussed in section IV-E.

\par To evaluate our method, nine widely used change detection methods are adopted for comparison, they are summarized as follows. 

\begin{enumerate} 
  \item IRMAD \cite{Nielsen2007}, which is an iteratively weighted extension of MAD and has shown good performance in multi-temporal image change detection.
  \item ISFA \cite{Wu2014}, which is an unsupervised change detection approach based on slow feature analysis theory.
  \item CVA \cite{Sharma2007}, which is one of the most classic unsupervised change detection methods.
  \item OBCD \cite{Desclee2006}, an unsupervised change detection method for VHR images, which adopts the object as the change detection basic unit.
  \item LSTM \cite{Lyu2016}, a deep learning-based method, has recently shown promising performance in change detection.
  \item PCANet \cite{Gao2016}, which utilizes Gabor wavelets and FCM as the pre-classification method to select training samples, and then trains a PCANet [48] model with the selected image patches.
  \item DSFA \cite{Du2019a}, a deep learning-based change detection algorithm, which works by extracting nonlinear features from multi-temporal images with two-stream DNN and detecting changes with SFA.
  \item DCVA \cite{Saha2019}, an effective unsupervised method for binary and multi-class change detection in VHR images, which adopts a pre-trained CNN extract deep spatial-spectral features from multi-temporal VHR images.
  \item DSCN \cite{Zhan2017}, a deep learning-based change detection method, which uses a deep siamese convolutional network to detect changes and performs well in aerial images.
\end{enumerate}

\par Among these methods, IRMAD, ISFA, CVA, OBCD, and DCVA are unsupervised models without training samples. PCANet and DSFA are pre-classification-based methods. LSTM and DSCN are supervised models, we train them on the same samples as the DSMS-CN. Precision rate, recall rate, overall accuracy (OA), F1 score, and kappa coefficient (KC) are used for accuracy assessment. 

\begin{figure}[t]
    
  \centering

  \subfloat[]{
    \includegraphics[width=1.6in]{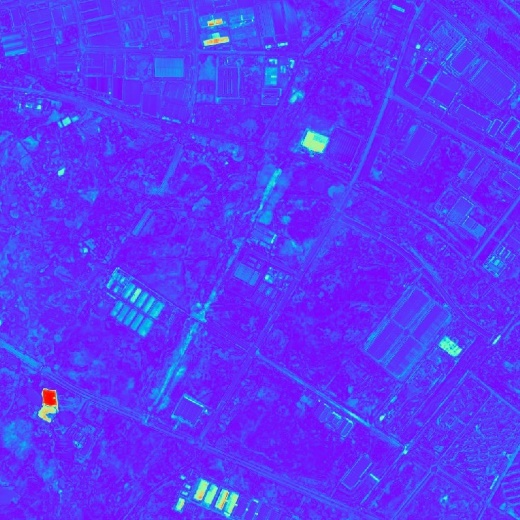}
  \label{fig_first_case}}
  \hfil
  \subfloat[]{
    \includegraphics[width=1.6in]{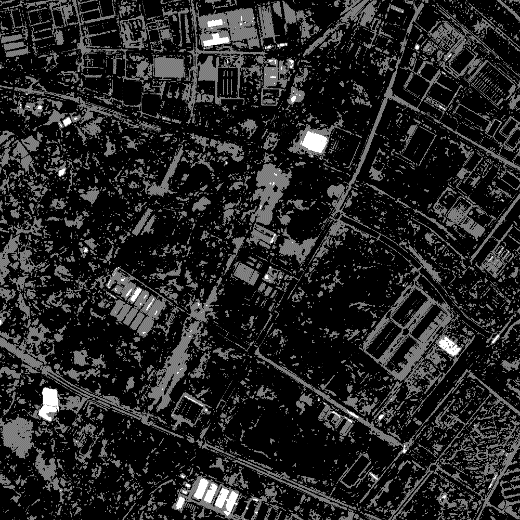}
  \label{fig_second_case}}

  \caption{Pre-classification results on the WH data set. (a) DI. (b) CICM.}
  \label{WH_prec}
\end{figure}

\begin{figure*}[t]
  \centering
  \subfloat[]{
    \includegraphics[width=1.3in]{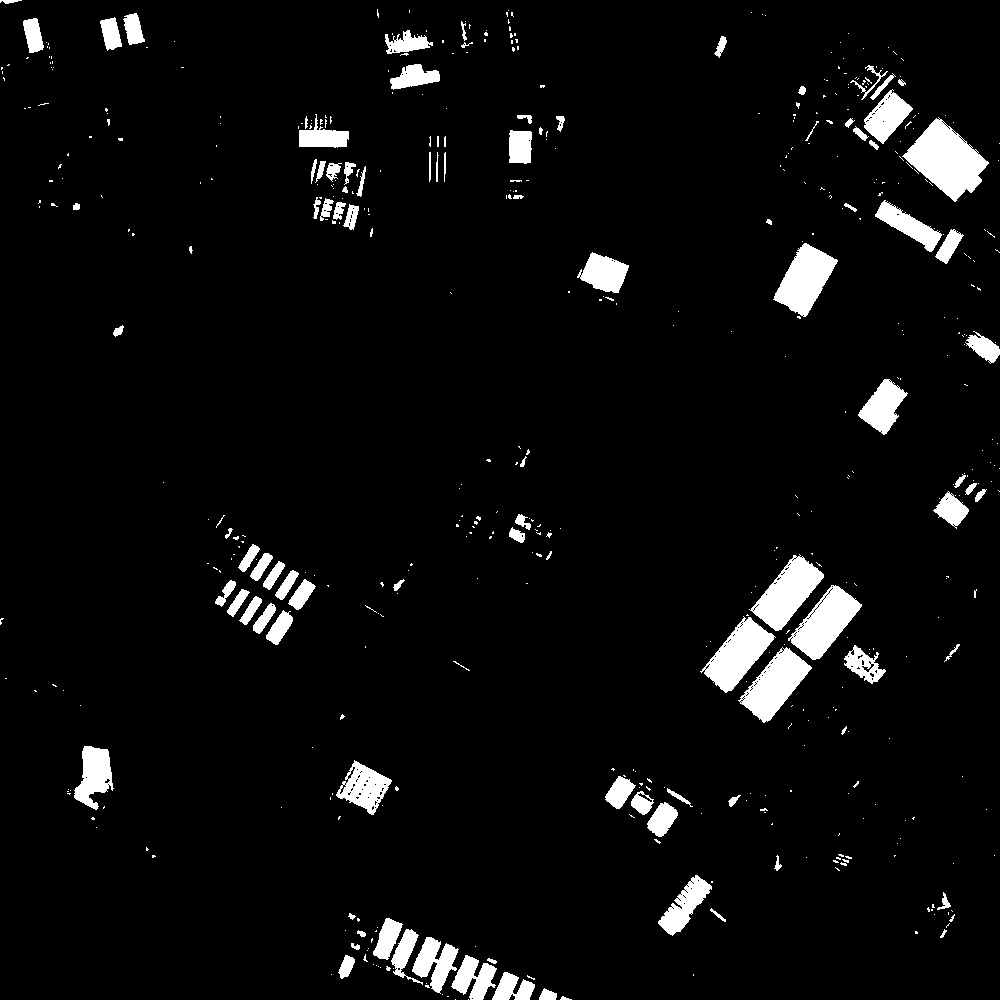}
  \label{fig_first_case}}
  \hfil
  \subfloat[]{
    \includegraphics[width=1.3in]{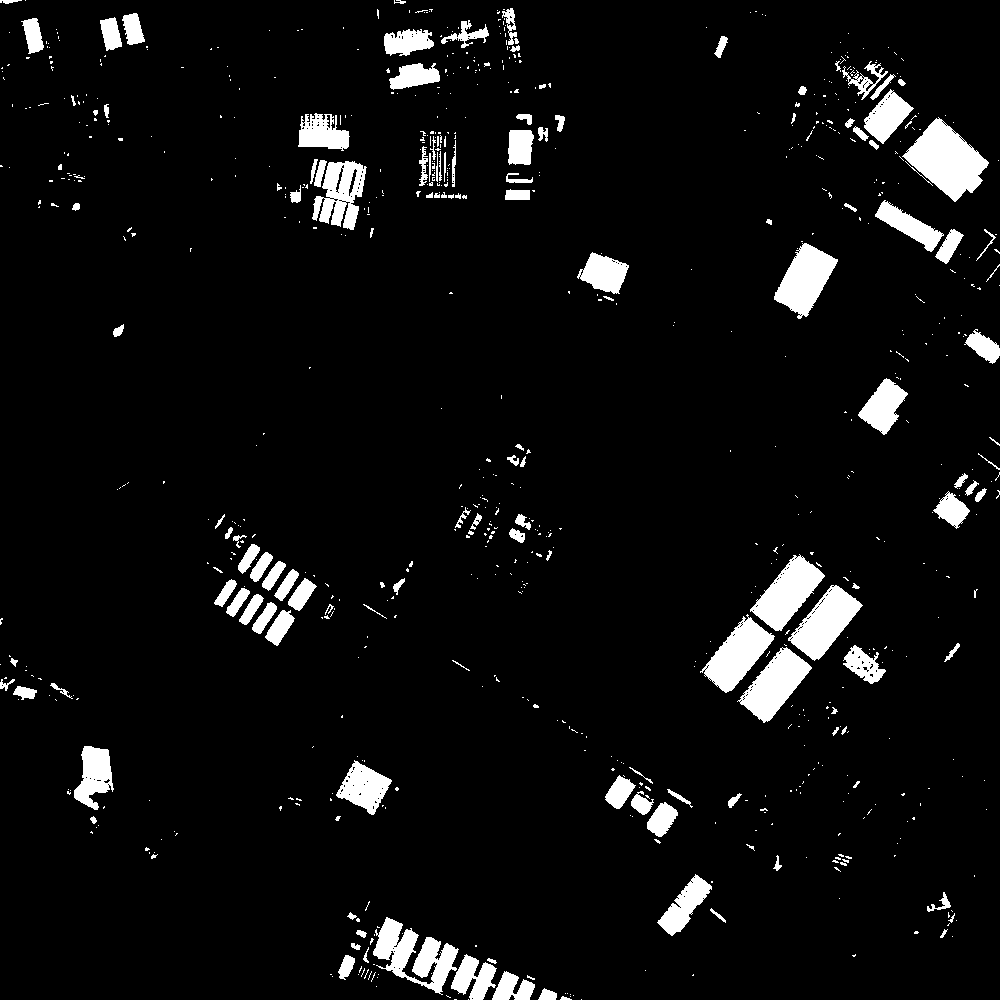}
  \label{fig_second_case}}
  \hfil
  \subfloat[]{
    \includegraphics[width=1.3in]{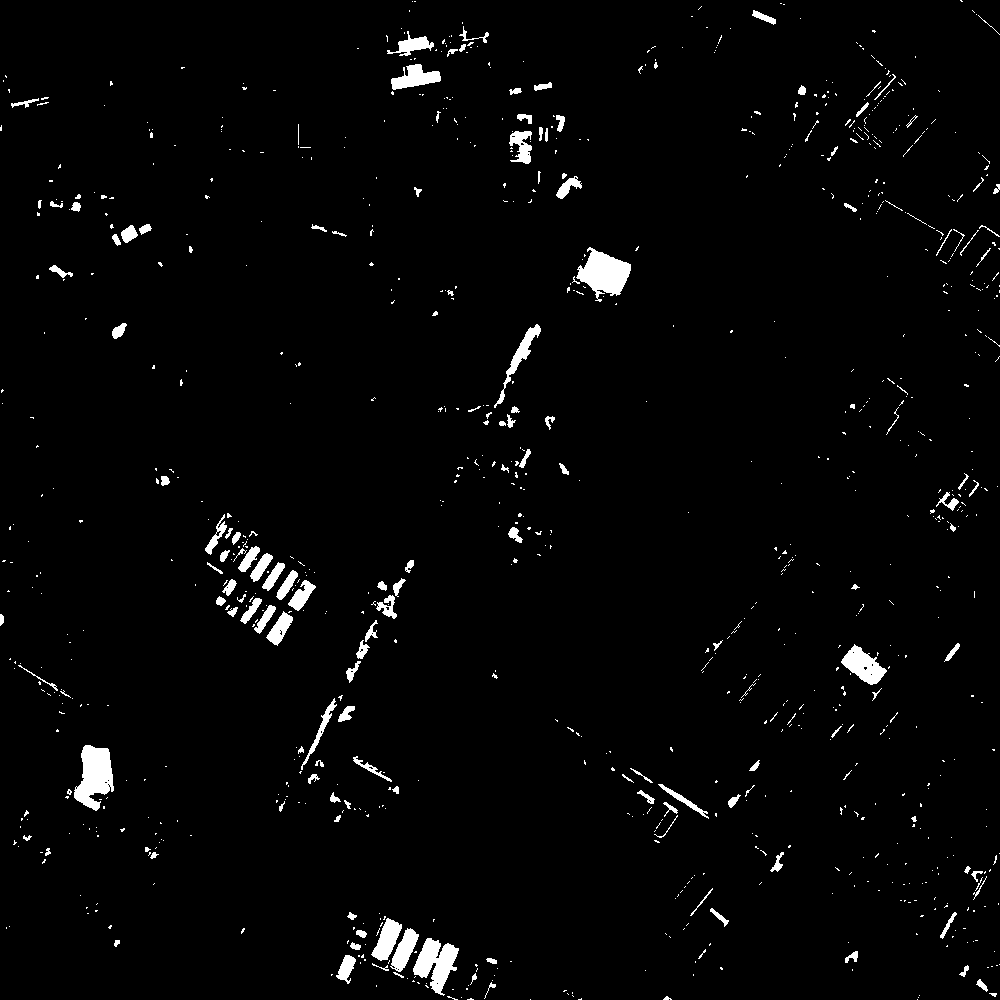}
  \label{fig_third_case}}
  \hfil
  \subfloat[]{
    \includegraphics[width=1.3in]{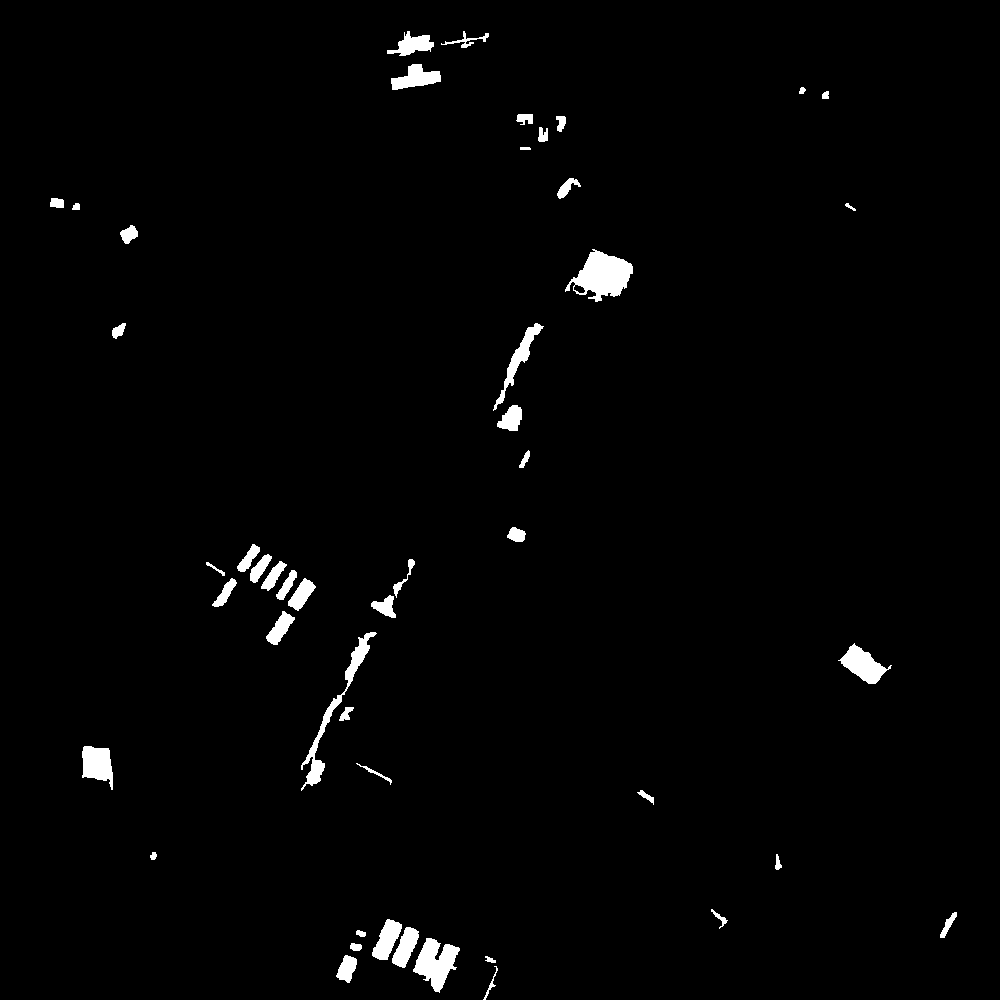}
  \label{fig_fourth_case}}
  \hfil
  \subfloat[]{
    \includegraphics[width=1.3in]{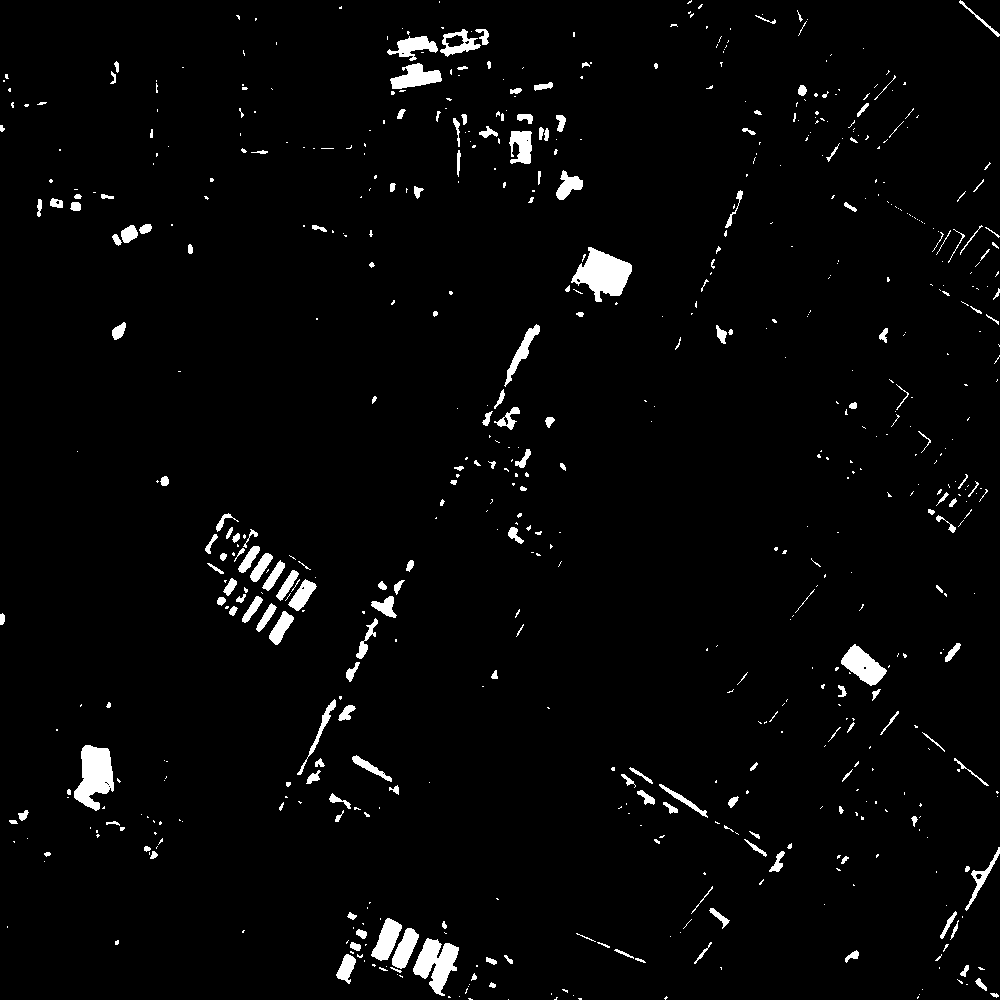}
  \label{fig_fifth_case}}
  
  \subfloat[]{
    \includegraphics[width=1.3in]{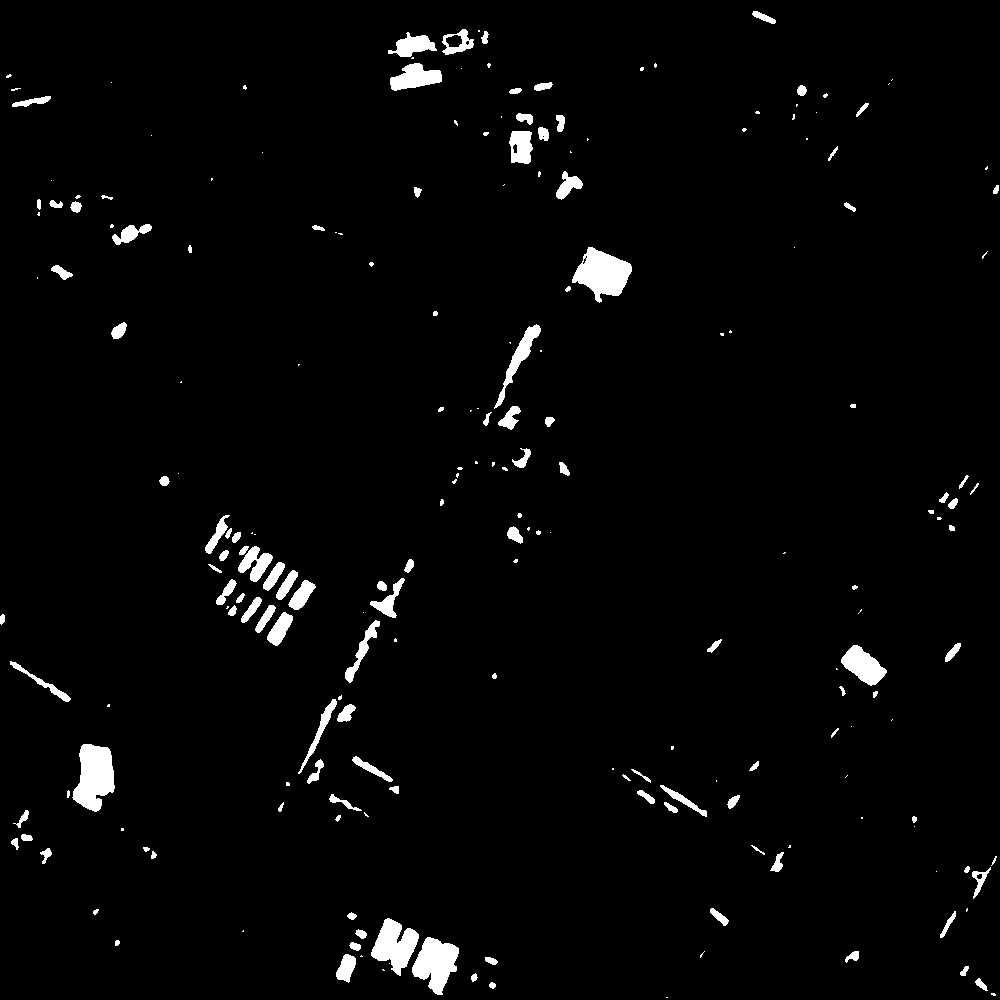}
  \label{fig_sixth_case}}
  \hfil
  \subfloat[]{
    \includegraphics[width=1.3in]{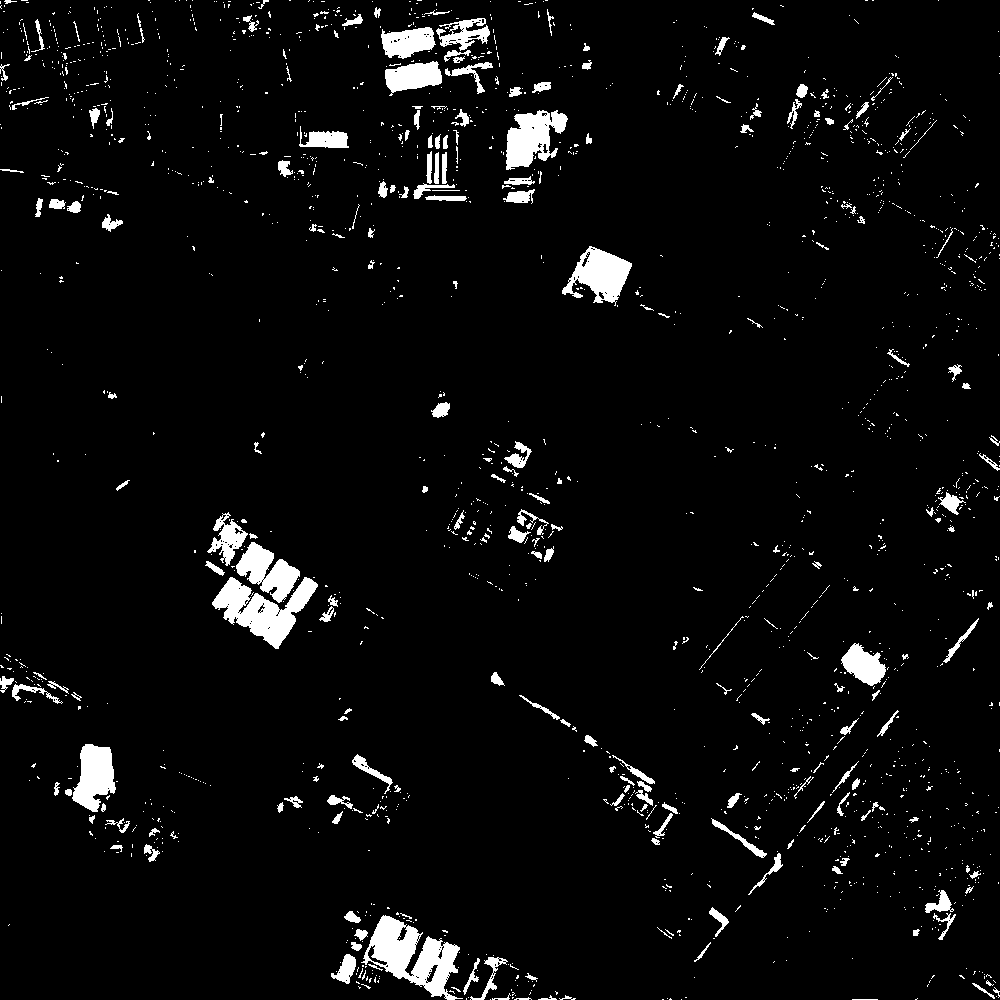}
  \label{fig_sixth_case}}
  \hfil
  \subfloat[]{
    \includegraphics[width=1.3in]{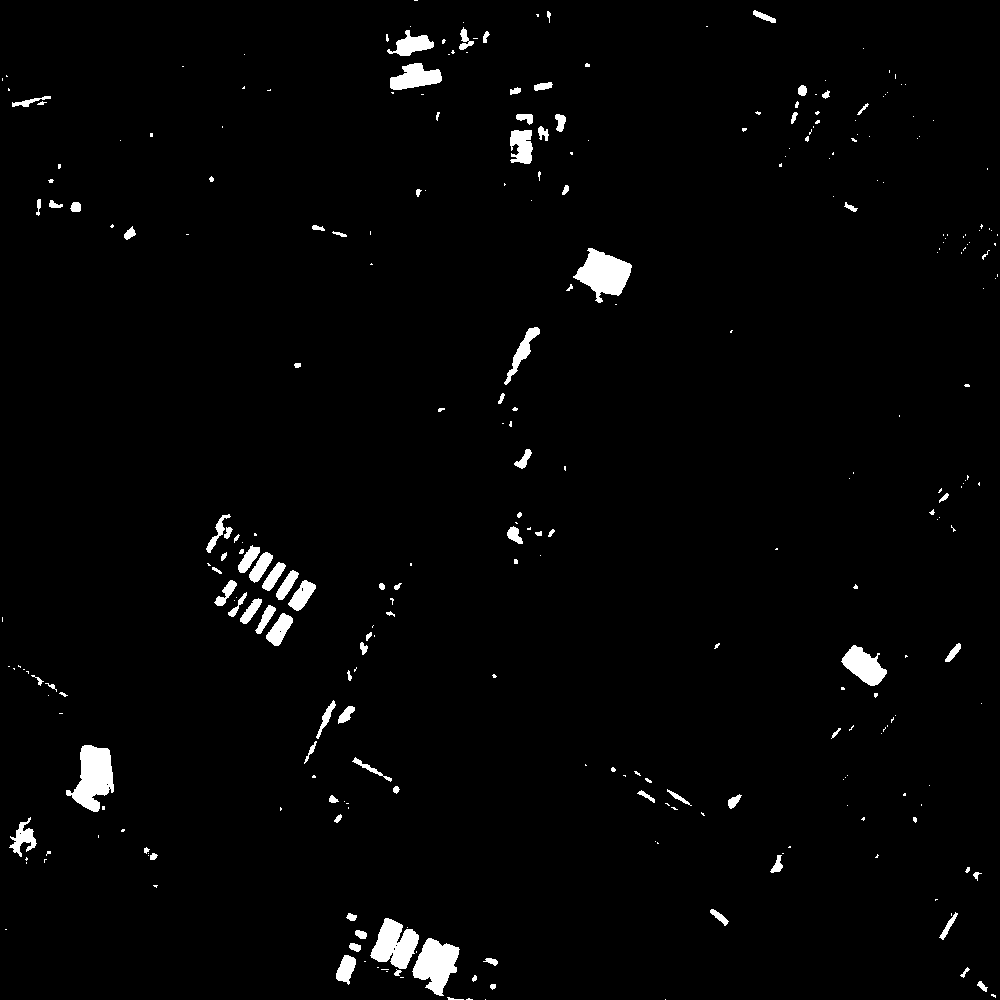}
  \label{fig_sixth_case}}
  \hfil
  \subfloat[]{
    \includegraphics[width=1.3in]{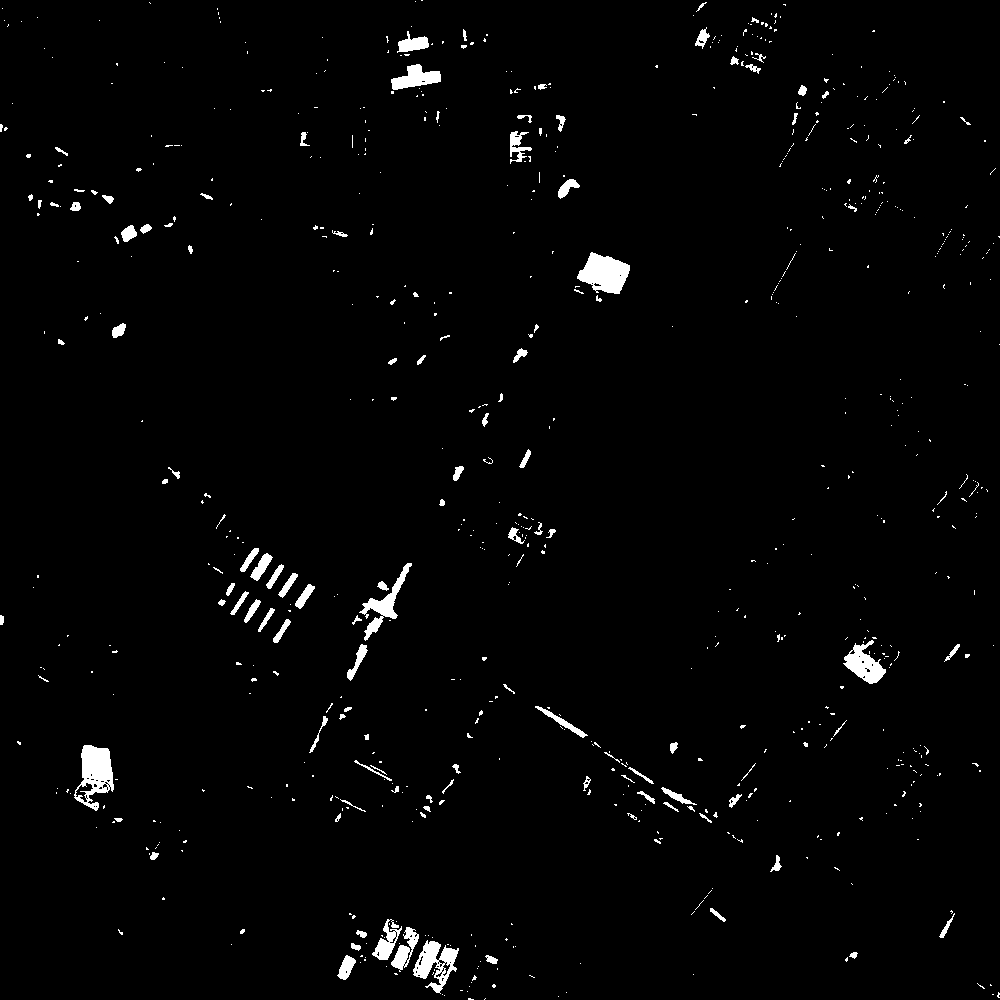}
  \label{fig_seventh_case}}
  \hfil
  \subfloat[]{
    \includegraphics[width=1.3in]{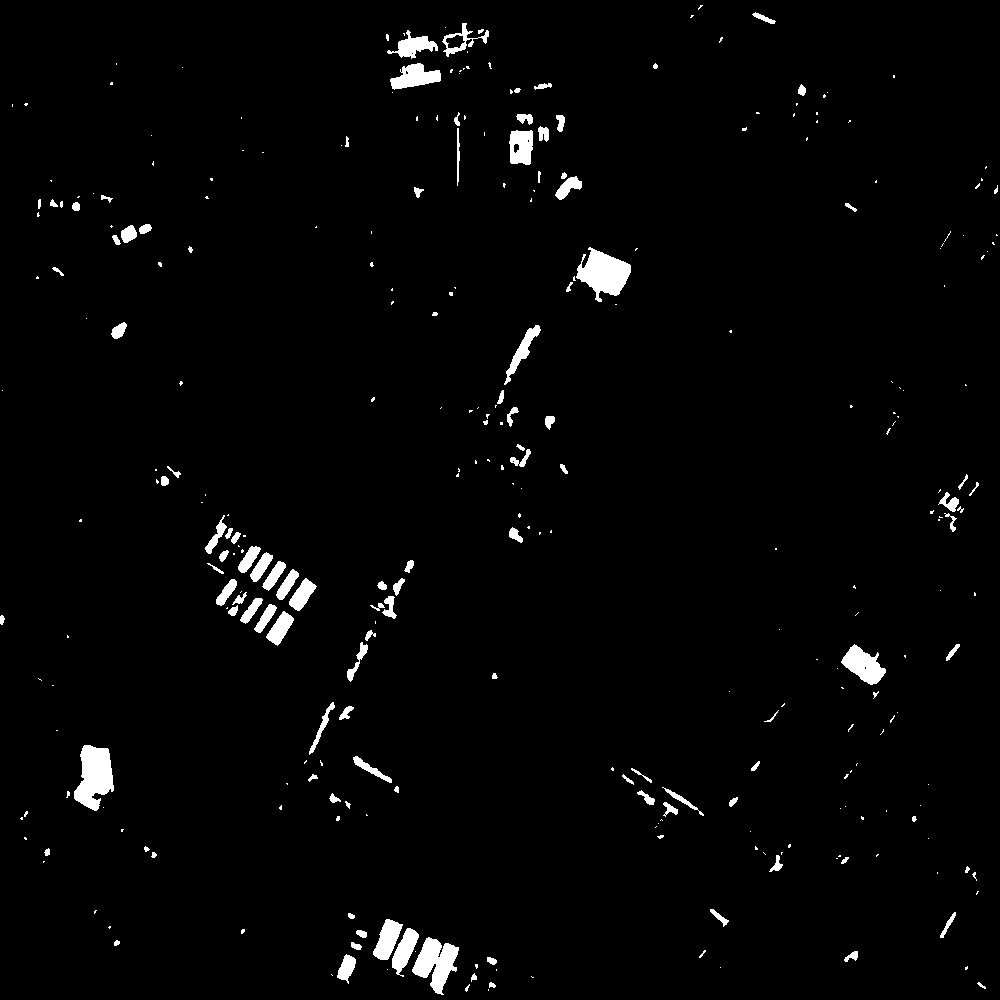}
  \label{fig_eighth_case}}
  \caption{Change detection results obtained by different methods on the WH data set. (a) IRMAD. (b) ISFA. (c) CVA. (d) OBCD. (e) LSTM. (f) PCANet. (g) DSFA. (h) DCVA. (i) DSCN. (j) DSMS-CN.}
  \label{WH_result}
\end{figure*}

\subsection{Experimental Result and Analysis on WH Data Set}
\par According to our unsupervised architecture, the WH data set is first pre-processed. Then, CVA fuses information of four bands and computes DI. As shown in Fig. \ref{WH_prec}-(a), the areas of warmer tones have a greater change probability. Based on DI, FCM is implemented to generate coarse initial change map (CICM). In Fig. \ref{WH_prec}-(b), the white pixels belong to change class, the dark pixels belong to non-change class, and gray pixels are candidates to be classified by the proposed method. 

\par All the pixels in the change class are chosen as training samples. However, since spectral difference between multi-temporal images of the pixels in non-change class is relatively constant, we only randomly select a part of pixels in the non-change class as training samples. On the WH data set, the number of selected unchanged pixels is four times to the changed ones. Section IV-E further discusses the impact of the proportion of change and non-change classes in the training sample.

\par The binary change maps obtained by DSMS-CN and comparison methods are presented in Fig. \ref{WH_result}. The change detection results acquired by IRMAD and ISFA are unsatisfactory, it is obvious that a lot of unchanged buildings are detected as nto changes, which indicates ISFA and IRMAD suffer from the “over-exposed” problem. Compared with IRMAD and ISFA, CVA directly calculating a DI and performing clustering achieves a relatively good result. However, changes of a few buildings roads are not recognized and many pixels in building margins are falsely detected as changed ones. Due to adopting object as the basic unit, there is almost no noise in the result of OBCD. But limited to only exploring low-level spatial-spectral features, lots of building changes are not detected. Fig. \ref{WH_result}-(e) shows the result of LSTM, there exist some noises caused by insufficient utilization of spatial context information. The result of PCANet is shown in Fig. \ref{WH_result}-(f), through cascade PCA filters, the noises are well suppressed. Extracting nonlinear spatial-spectral features by DNN, most of the changes are well preserved by DSFA as shown in Fig. \ref{WH_result}-(g). However, because of detecting changes through SFA, DSFA is inevitably affected by the “over-exposed” problem. Through comparing the deep spatial-spectral features extracted from input images, DCVA shows a good change detection performance though a part of changes is misclassified. From Fig. \ref{WH_result}-(i), it can be observed that plenty of building changes are not detected by DSCN. Compared with these methods, the binary change map of the proposed DSMS-CN is better in visual, as shown in Fig. \ref{WH_result}-(j).

\begin{table}[t]
  \captionsetup{font={small}}
  \renewcommand{\arraystretch}{1.3}
  \caption{ACCURACY ASSESSMENT ON THE BINARY CHANGE MAPS ACQUIRED BY DIFFERENT METHODS ON THE WH DATA SET}
  \label{WH_table}
  \centering
  \begin{tabular}{c c c c c c}
    \hline
    \bfseries Method & \bfseries Pre. & \bfseries Rec.  & \bfseries OA  & \bfseries F1 & \bfseries KC \\
    \hline\hline
    IRMAD	& 0.3651 & 0.5708	& 0.9212	& 0.3651	& 0.3289 \\ 	
    ISFA	& 0.2790	& 0.6408	& 0.9200	& 0.3888	& 0.3530 \\  									
    CVA	& 0.7682		& 0.5679	& 0.9711	& 0.6530	& 0.6434 \\ 									
    OBCD & \textbf{0.9409}	& 0.4906	& 0.9785	& 0.6449	& 0.6350 \\  								
    LSTM	& 0.7761	& 0.6125	& 0.9782	& 0.6892	& 0.6780 \\ 	 							
    PCANet	& 0.8386	& 0.6282	& 0.9805	& 0.7183	& 0.7084 \\
    DSFA	& 0.6936	& \textbf{0.7770}	& 0.9776	& 0.7329	& 0.7212 \\	
    DCVA	& \underline{0.8665}	& 0.6196	& \underline{0.9812}	& 0.7225	& 0.7131 \\						
    DSCN	& 0.7852	& 0.4484	& 0.9729	& 0.5708	& 0.5564 \\ 		 							
    DSMS-CN	& 0.8120	& \underline{0.6844} & \textbf{0.9813}	& \textbf{0.7428} & \textbf{0.7331} \\ 									
    \hline
  \end{tabular}
\end{table}

\par Table \ref{WH_table} reports the quantitative analysis results based on five evaluation criteria as described in Section IV-B. DSMS-CN achieves the best result with OA of 0.9812, F1 of 0.7428, and KC of 0.7331. It indicates that DSMS-CN can effectively fit the distributions of ground changes in VHR images from pre-classification samples based on the powerful extraction ability of MFCU and deep siamese convolutional structure, and achieve the best performance. By contrast, the performance of SVM and DSCN cannot compete with our approach.

\begin{figure}[t]
    
  \centering

  \subfloat[]{
    \includegraphics[width=1.6in]{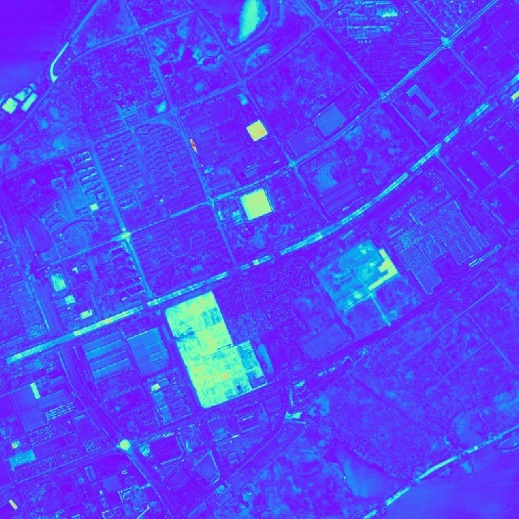}
  \label{fig_first_case}}
  \hfil
  \subfloat[]{
    \includegraphics[width=1.6in]{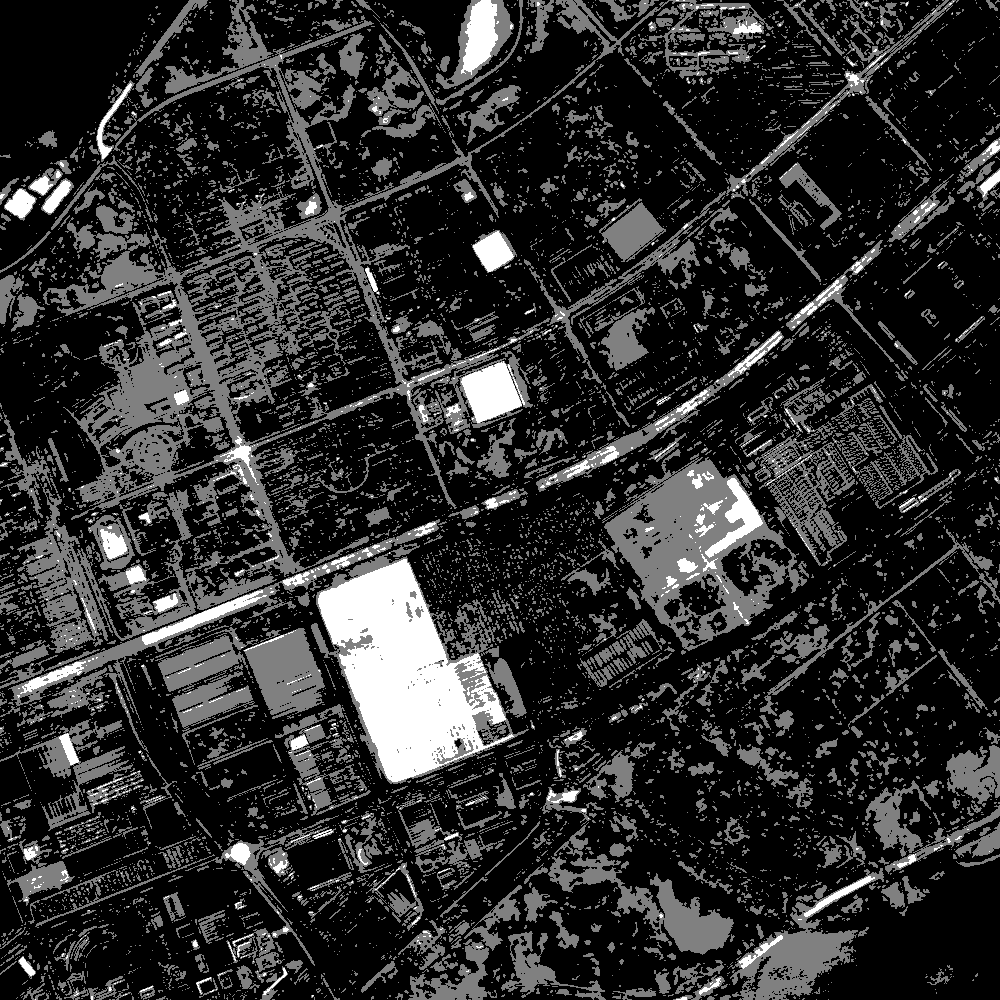}
  \label{fig_second_case}}

  \caption{Pre-classification results on the HY data set. (a) DI. (b) CICM.}
  \label{HY_prec}
\end{figure}

\begin{figure*}[ht]
  \centering
  \subfloat[]{
    \includegraphics[width=1.3in]{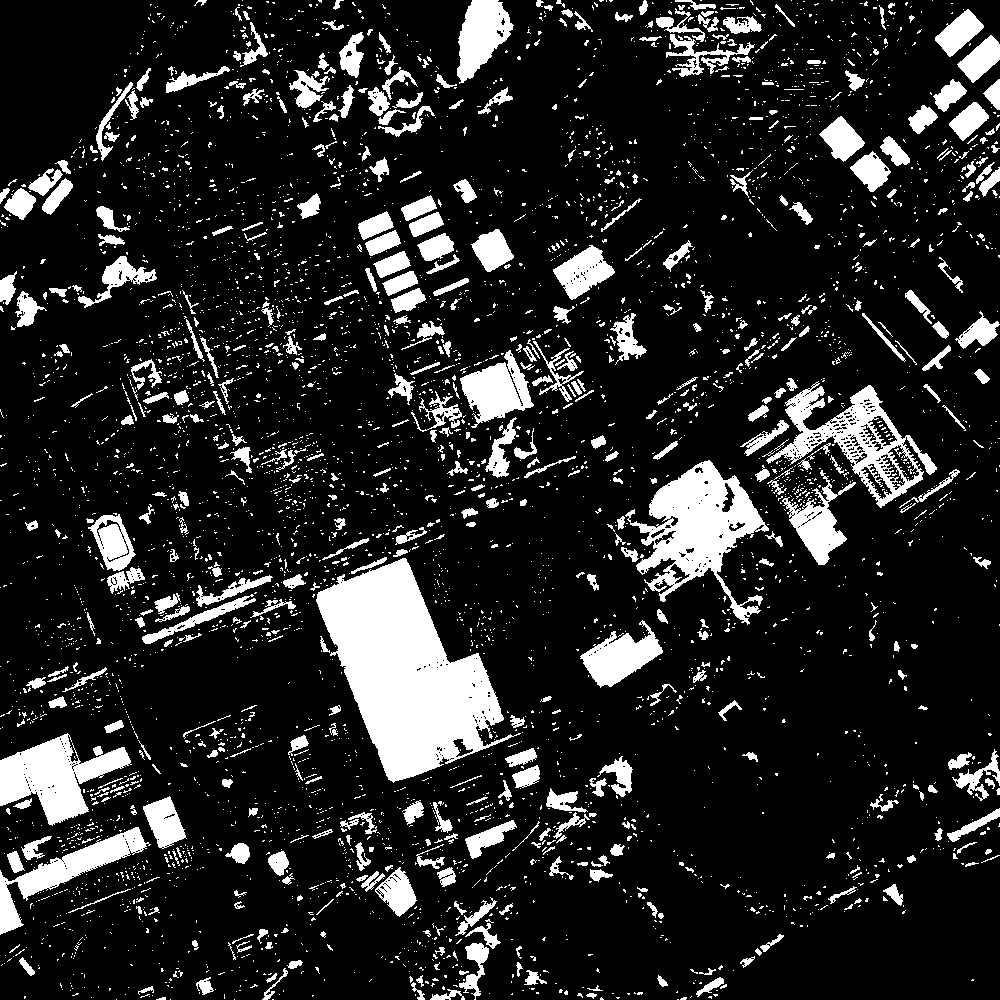}
  \label{fig_first_case}}
  \hfil
  \subfloat[]{
    \includegraphics[width=1.3in]{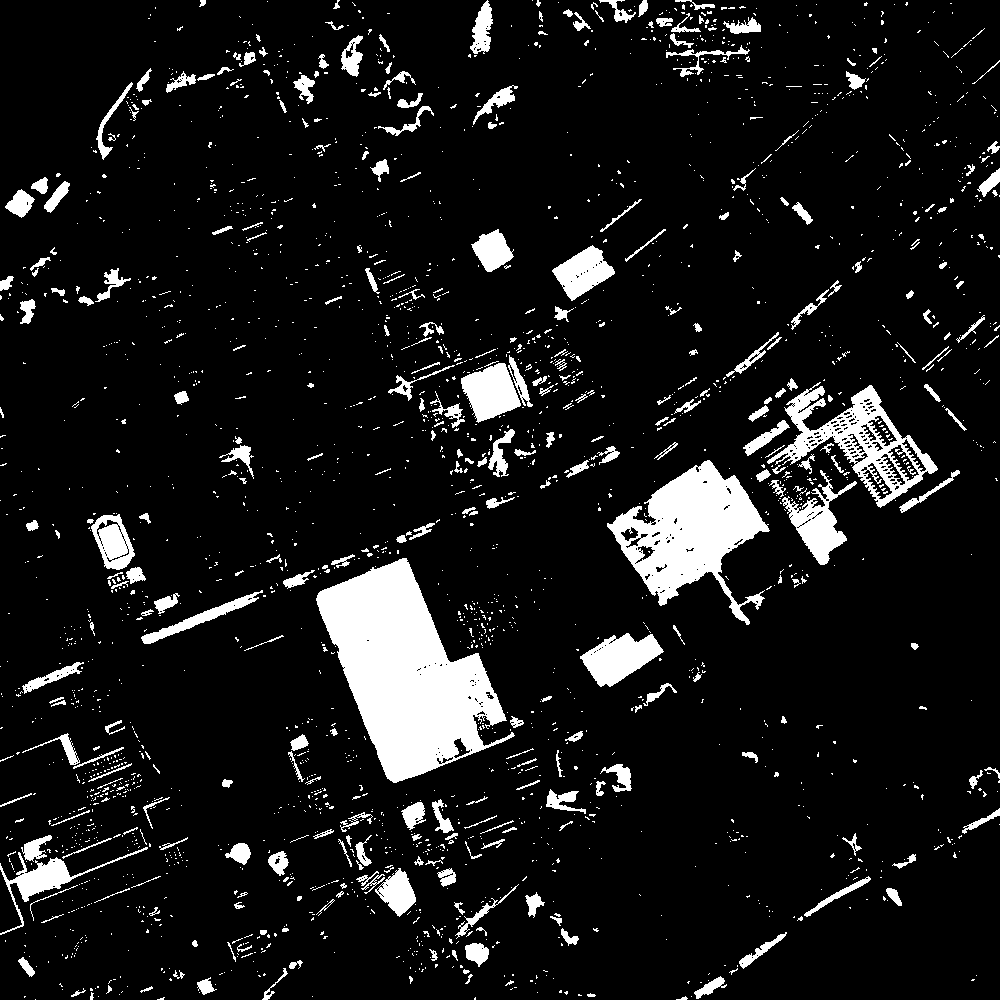}
  \label{fig_second_case}}
  \hfil
  \subfloat[]{
    \includegraphics[width=1.3in]{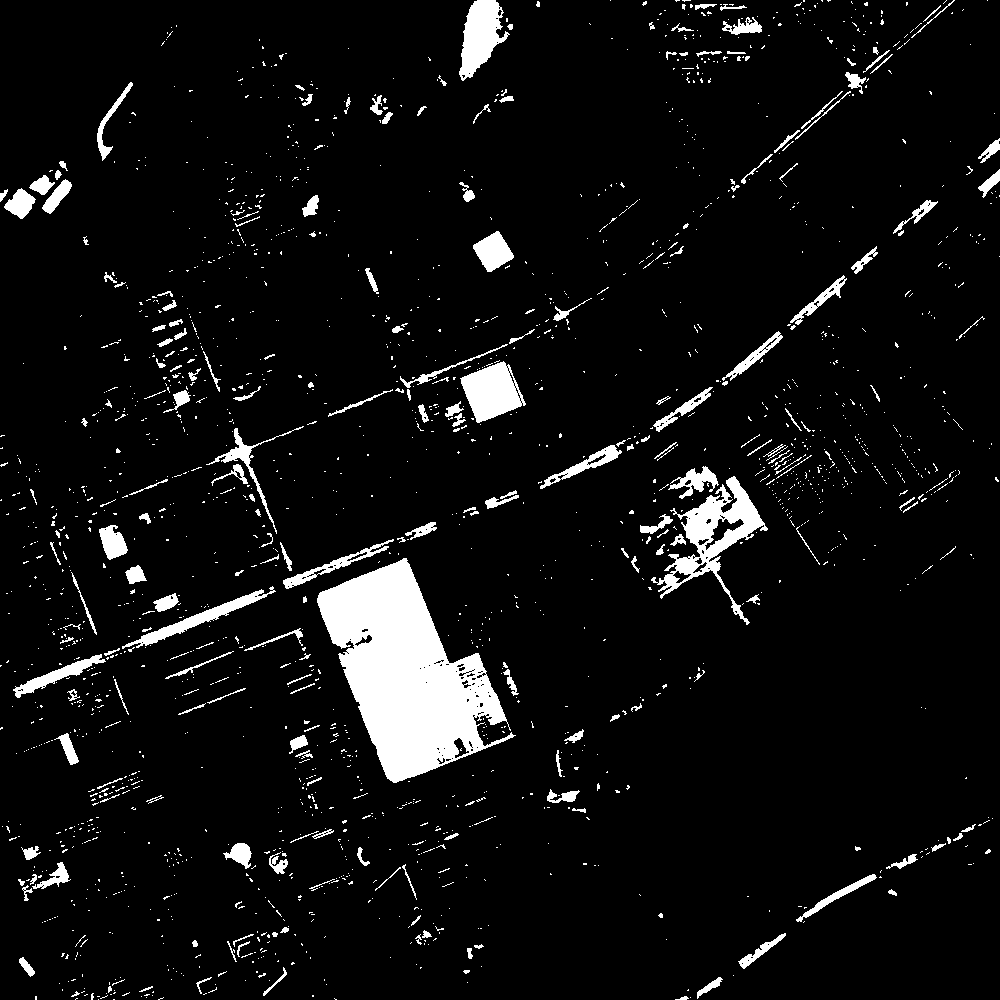}
  \label{fig_third_case}}
  \hfil
  \subfloat[]{
    \includegraphics[width=1.3in]{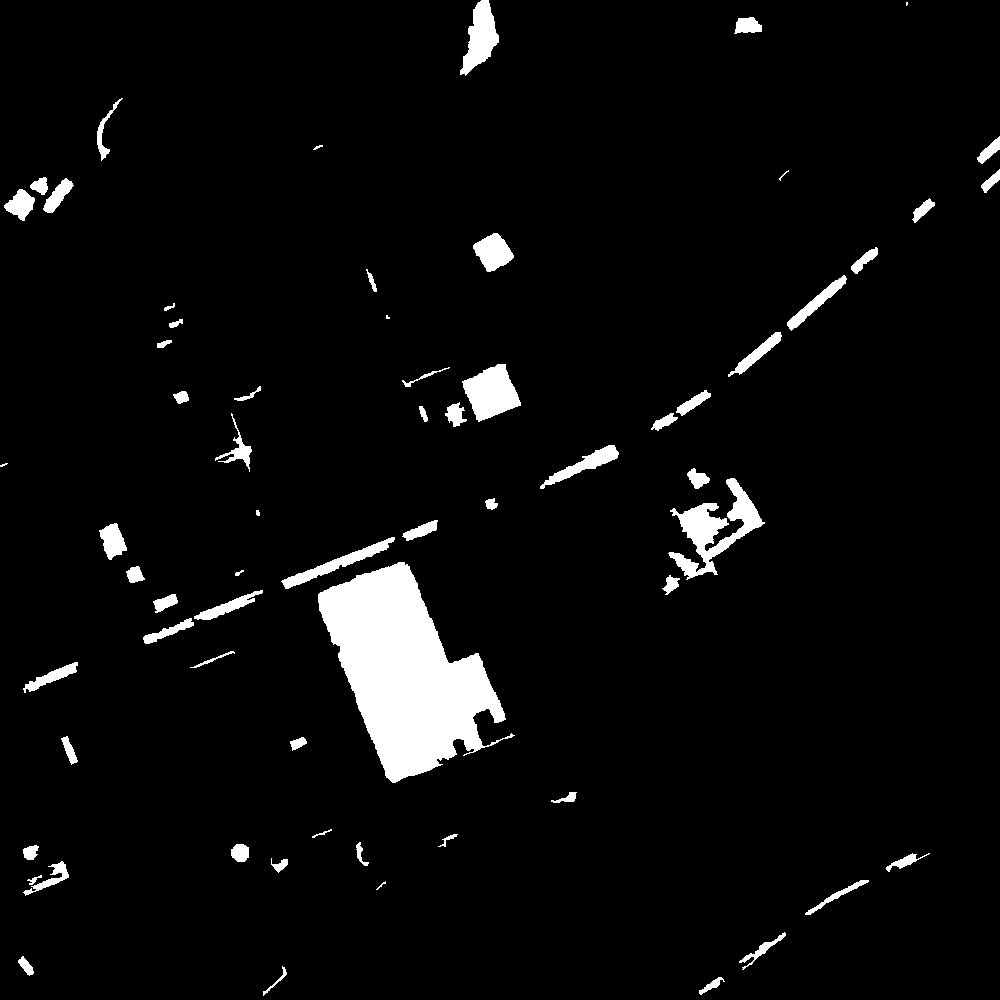}
  \label{fig_fourth_case}}
  \hfil
  \subfloat[]{
    \includegraphics[width=1.3in]{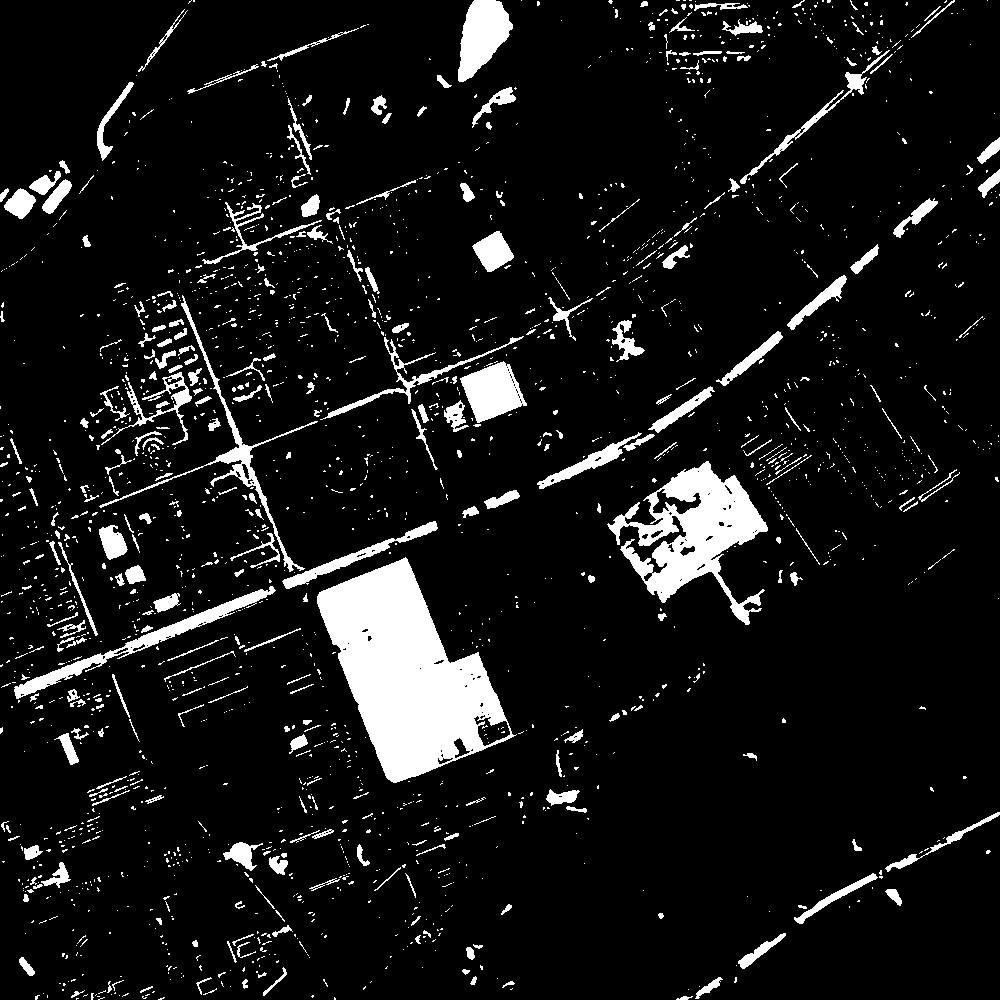}
  \label{fig_fifth_case}}
  
  \subfloat[]{
    \includegraphics[width=1.3in]{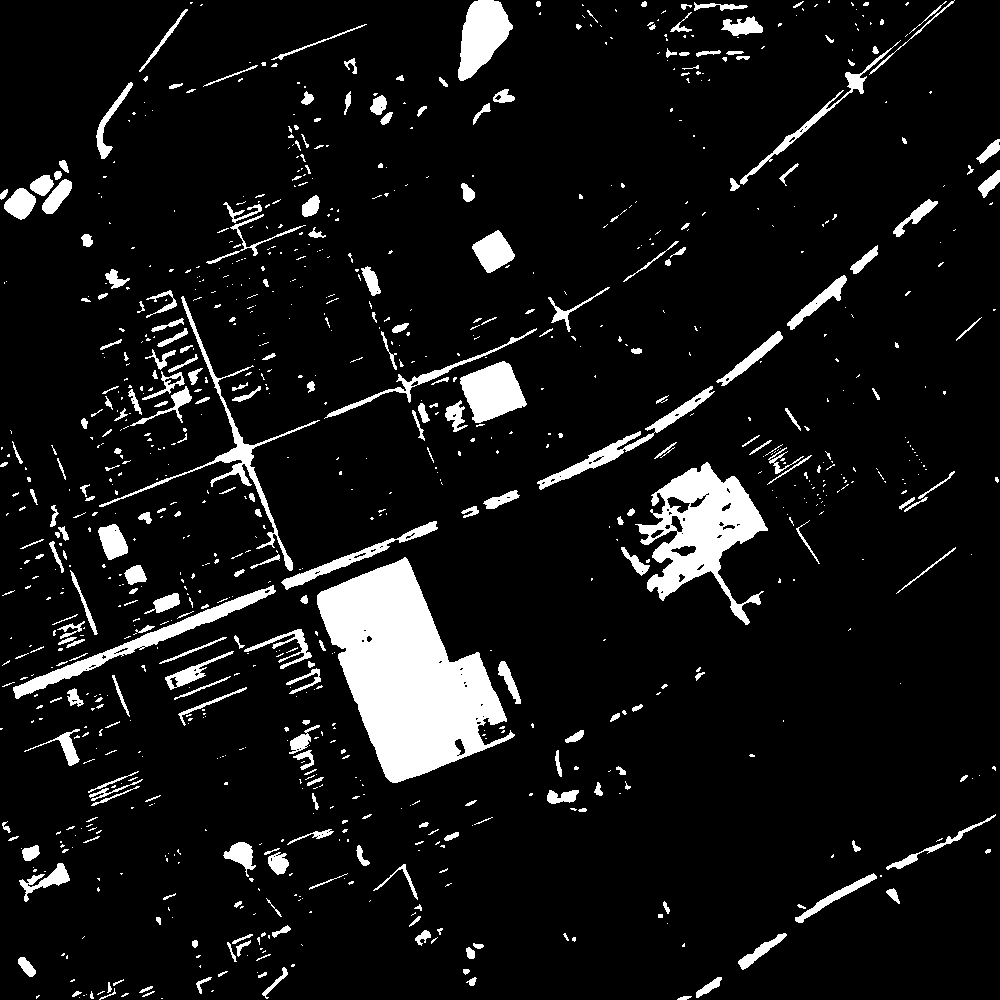}
  \label{fig_sixth_case}}
  \hfil
  \subfloat[]{
    \includegraphics[width=1.3in]{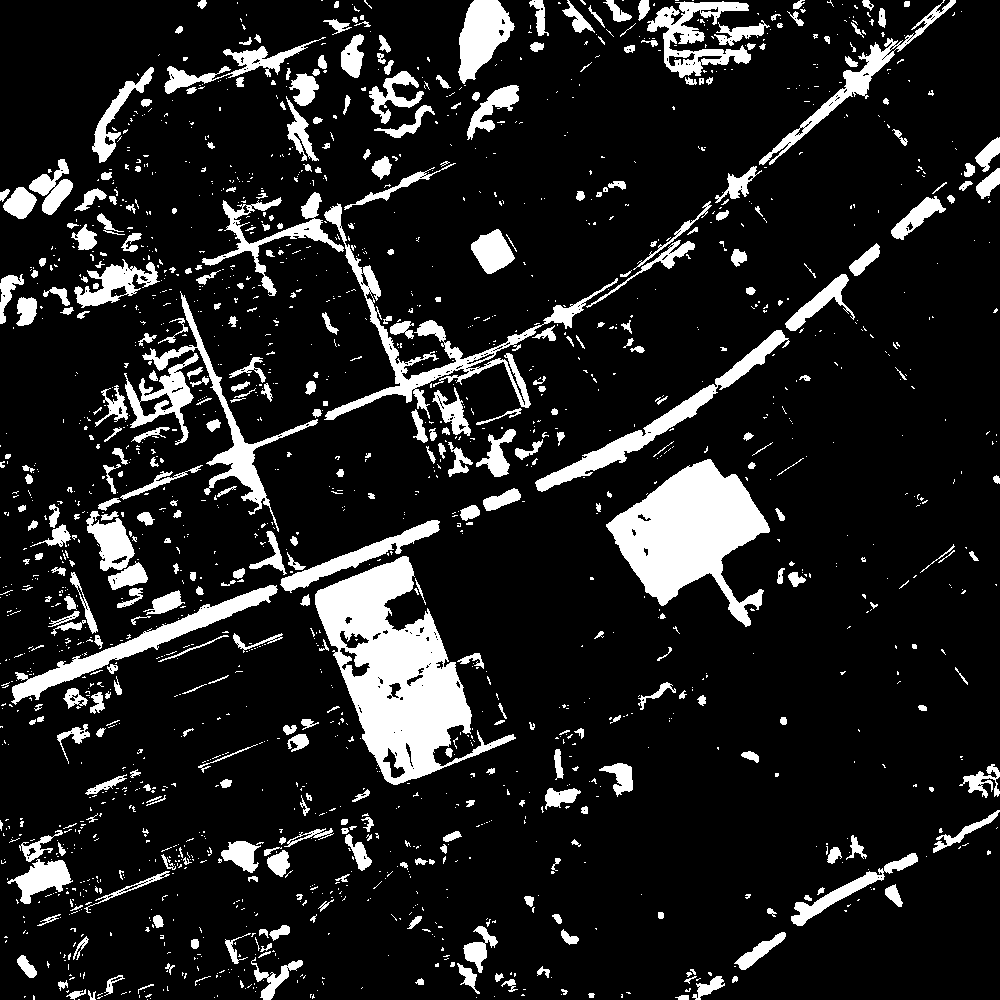}
  \label{fig_sixth_case}}
  \hfil
  \subfloat[]{
    \includegraphics[width=1.3in]{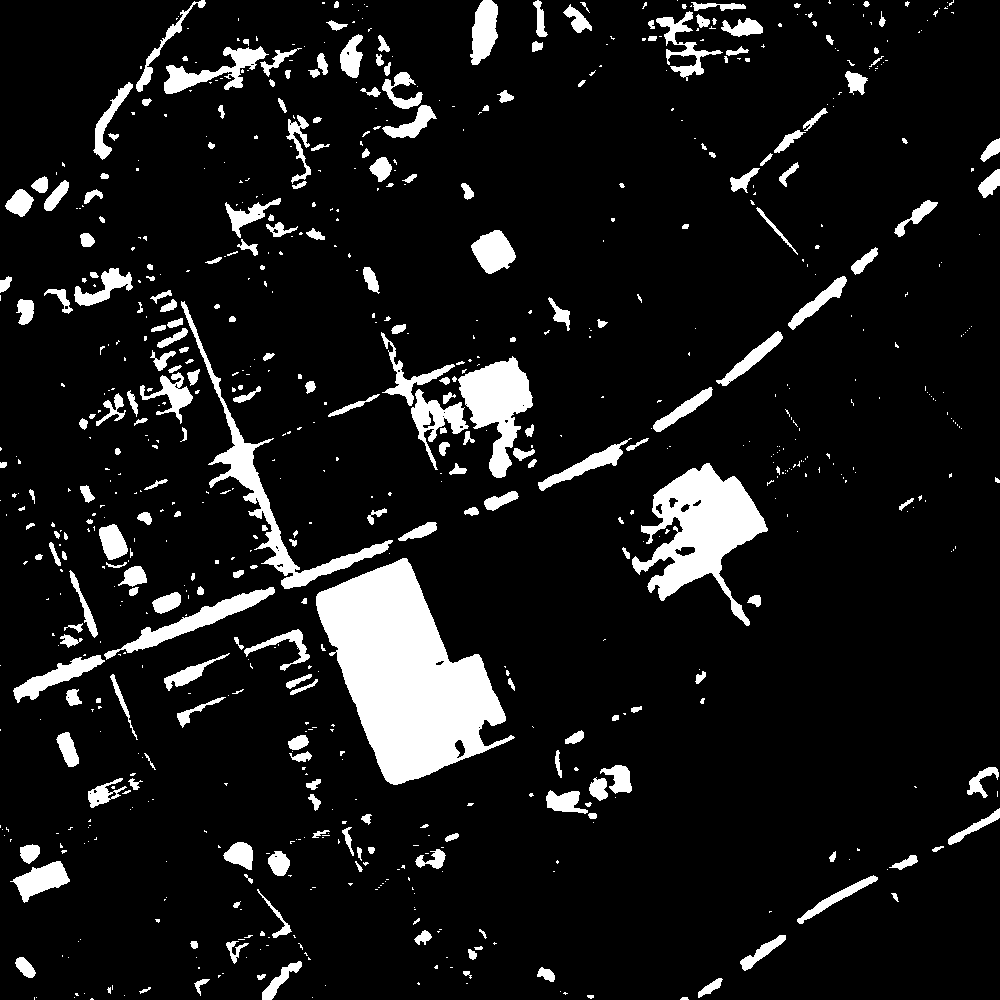}
  \label{fig_sixth_case}}
  \hfil
  \subfloat[]{
    \includegraphics[width=1.3in]{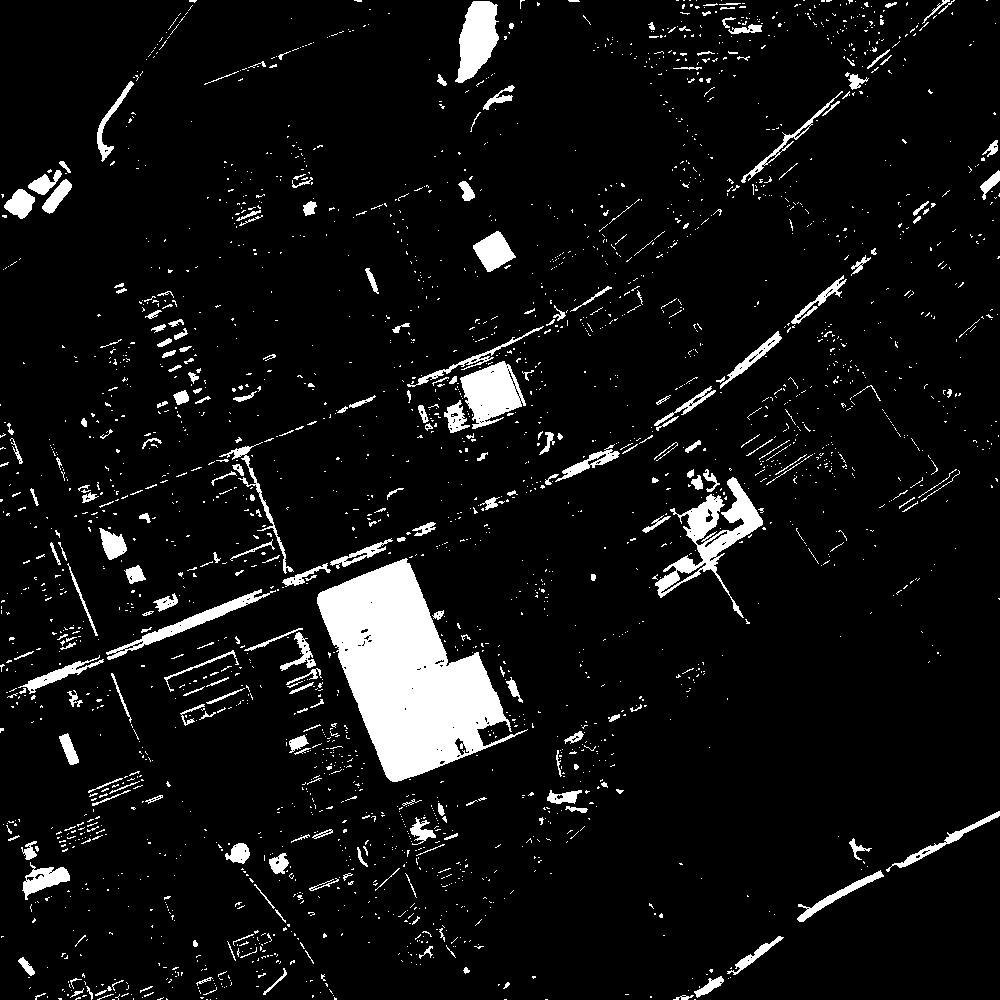}
  \label{fig_seventh_case}}
  \hfil
  \subfloat[]{
    \includegraphics[width=1.3in]{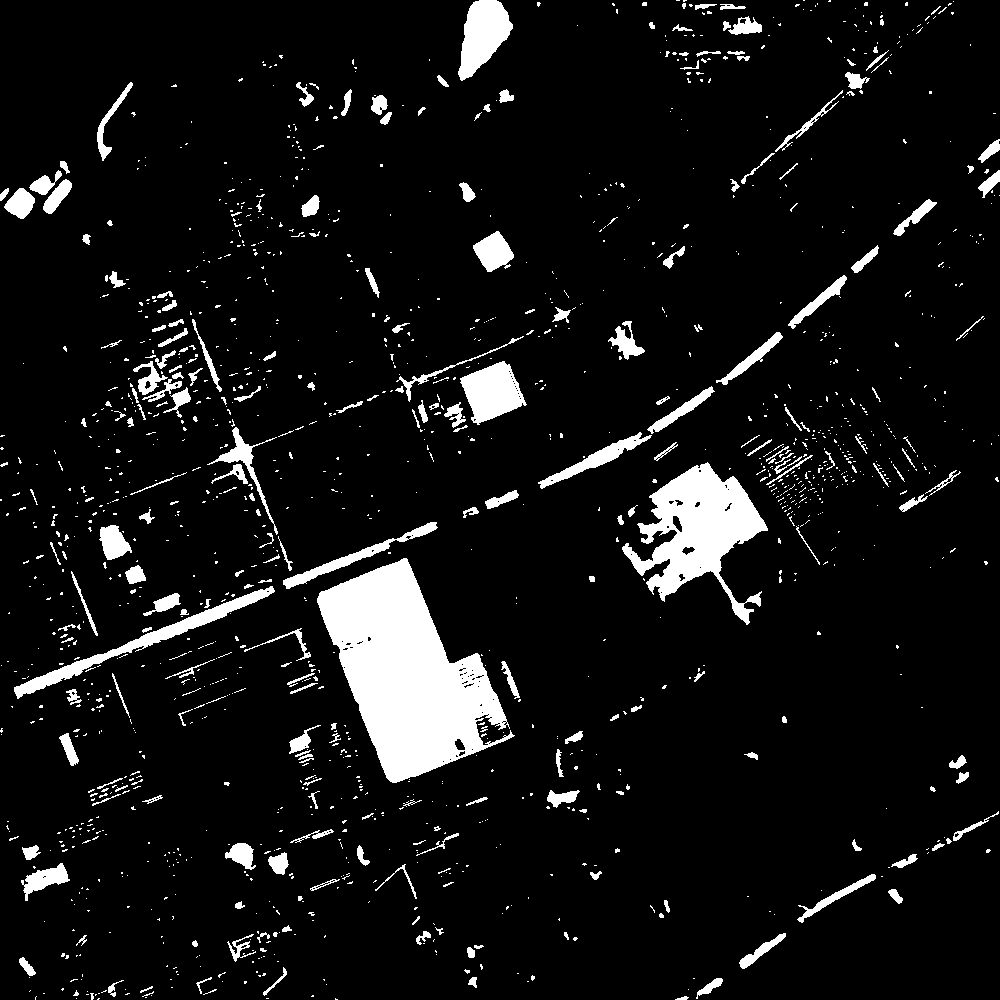}
  \label{fig_eighth_case}}
  \caption{Change detection results obtained by different methods on the HY data set. (a) IRMAD. (b) ISFA. (c) CVA. (d) OBCD. (e) LSTM. (f) PCANet. (g) DSFA. (h) DCVA. (i) DSCN. (j) DSMS-CN.}
  \label{HY_result}
\end{figure*}

\begin{table}[t]
  \captionsetup{font={small}}
  \renewcommand{\arraystretch}{1.3}
  \caption{ACCURACY ASSESSMENT ON THE BINARY CHANGE MAPS ACQUIRED BY DIFFERENT METHODS ON THE HY DATA SET}
  \label{HY_table}
  \centering
  \begin{tabular}{c c c c c c}
  \hline
  \bfseries Method & \bfseries Pre. & \bfseries Rec. & \bfseries OA & \bfseries F1 & \bfseries KC\\
  \hline\hline
  IRMAD	& 0.4356 & 0.7110	& 0.8497	& 0.5402	& 0.4565 \\ 	
  ISFA	& 0.6481	& 0.7120	& 0.9162	& 0.6786	& 0.6305 \\  									
  CVA	& 0.8579		& 0.6631	& 0.9445	& 0.7480	& 0.7174 \\ 									
  OBCD & \textbf{0.9802}	& 0.6161	& \underline{0.9507}	& 0.7566	& 0.7308 \\  								
  LSTM	& 0.7761	& 0.7680	& 0.9438	& 0.7720	& 0.7400 \\ 	 							
  PCANet	& 0.7891	& 0.7681	& 0.9458	& 0.7785	& 0.7476 \\
  DSFA	& 0.7432	& \textbf{0.8380}	& 0.9440	& 0.7877	& 0.7556 \\	
  DCVA	& 0.7928	& 0.7708	& 0.9466	& 0.7816	& 0.7512 \\						
  DSCN	& 0.7865	& 0.6428	& 0.9340	& 0.7074	& 0.6706 \\ 		 							
  DSMS-CN	& \underline{0.8304}	& \underline{0.7736} & \textbf{0.9523}	& \textbf{0.8011} & \textbf{0.7740} \\ 									
  \hline
  \end{tabular}
\end{table}

\subsection{Experimental Result and Analysis on HY Data Set}
\par The DI and CICM of the HY data set are shown in Fig. \ref{HY_prec}. Same as the WH data set, all the pixels in change class are chosen as training samples. But the number of selected unchanged pixels is the same as the changed ones. Section IV-E further discusses the impact of the proportion of change and non-change classes in the training sample.

\par The qualitative results obtained by the proposed method and comparison methods are shown in Fig. \ref{HY_result}. Fig. \ref{HY_result}-(a) and -(b) are the results obtained by IRMAD and ISFA. Because of the complex ground situations of VHR images, the results have many falsely detected pixels. Many unchanged buildings are misclassified as changes and the road changes are almost not detected. By contrast, the binary change maps obtained by CVA and OBCD are better in visual. However, CVA misclassifies a part of building margins as change class and OBCD ignores some obvious building and road changes. The result of LSTM is shown in Fig. \ref{HY_result}-(e). Through three gates and cell state to capture the temporal dependency, the main changes regions are detected successfully. But it still has plenty of noise. In Fig. \ref{HY_result}-(f), compared to LSTM, the binary change map of PCANet is better with less noise. As shown in Fig. \ref{HY_result}-(g), the road changes are well detected by DSFA, yet there is some internal fragmentation of changed regions. Fig. \ref{HY_result}-(h) shows that utilizing deep spatial-spectral features for change detection, the result generated by DCVA is obviously better than the result of CVA. The result obtained by DSCN is similar to the ones acquired by CVA, but more building margins are misclassified as change class. The change detection result by applying DSMS-CN on the HY data set is shown in Fig. \ref{HY_result}-(j). Compared with other methods, DSMS-CN is successful to identify most of the land-cover changes without losing many details and well preserve most of the unchanged regions, which further proves the powerful extraction ability of MFCU and effectiveness of the proposed network. Table \ref{HY_table} reporting the quantitative analysis results demonstrate it. Clearly, DSMS-CN achieves the highest OA, F1, and KC again.

\subsection{Discussion}
\begin{figure}[t]

  \centering
  \includegraphics[scale=0.5]{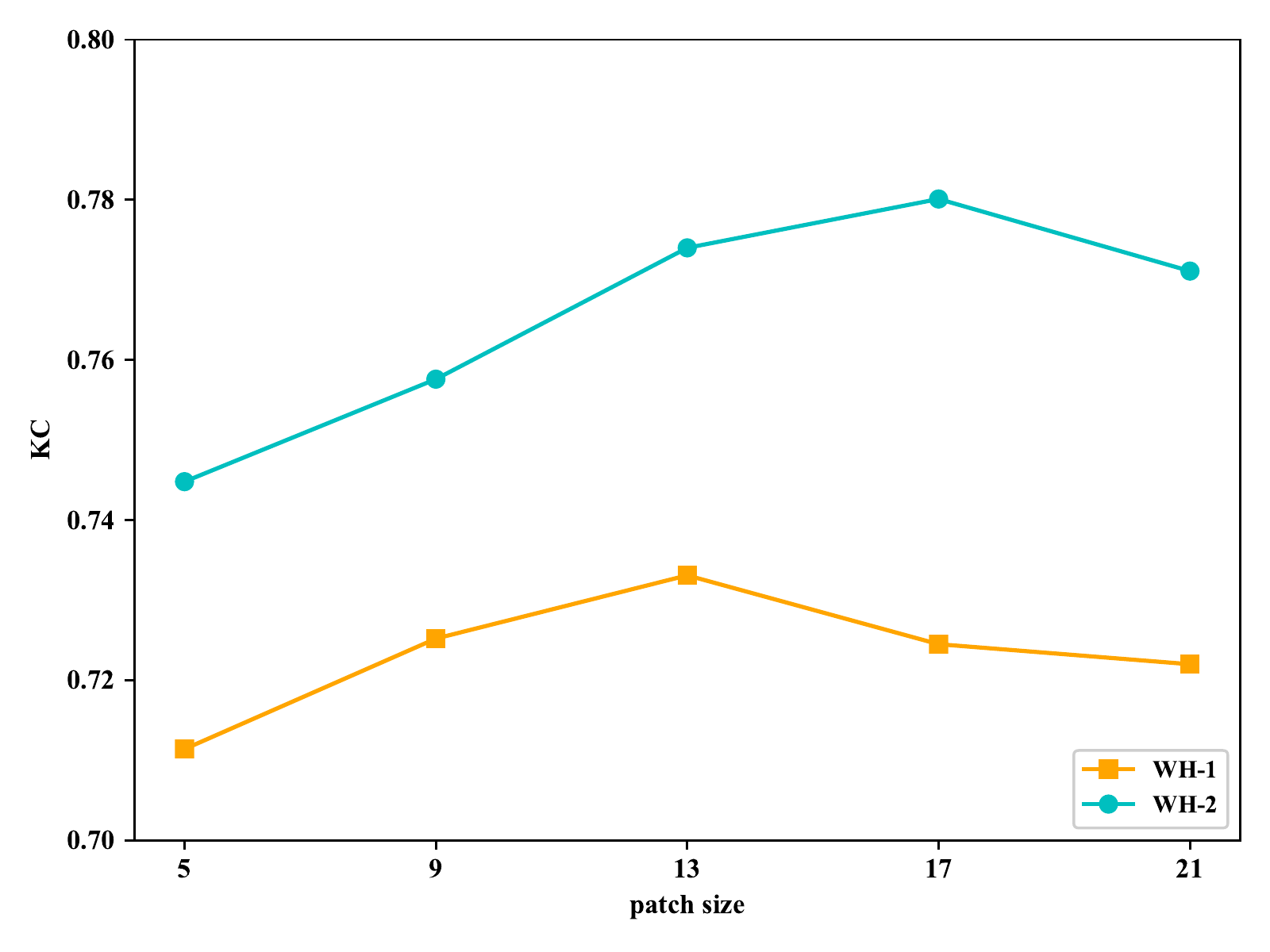}
  \caption{Relationship between the performance of the DSMS-CN and patch size.}
  \label{fig_patch_sz}
\end{figure}

\par For the proposed DSMS-CN, there are two hyper-parameters play an important role, namely input image patch size and proportion of non-change and change classes in training samples. The relationship between the change detection performance of the DSMS-CN and input image patch size is shown in Fig. \ref{fig_patch_sz}. It can be observed that the performance of DSMS-CN becomes better with the increasing of patch size on both data sets. This is because the larger the patch size is, the more neighborhood information the patch contains, which is beneficial for exploring multi-scale features by DSMS-CN to achieve better performance. However, if patch size is too large, more irrelevant interference information would be considered, resulting in impaired performance. Concretely speaking, the optimal values of patch size on the WH data set and HY data set are 13 and 17, respectively. But as the patch size increases, the computational expense would increase dramatically. Therefore, as a trade-off between performance and computational expense, the patch size is set to 13 on both data sets.

\begin{figure}[t]

  \centering
  \includegraphics[scale=0.5]{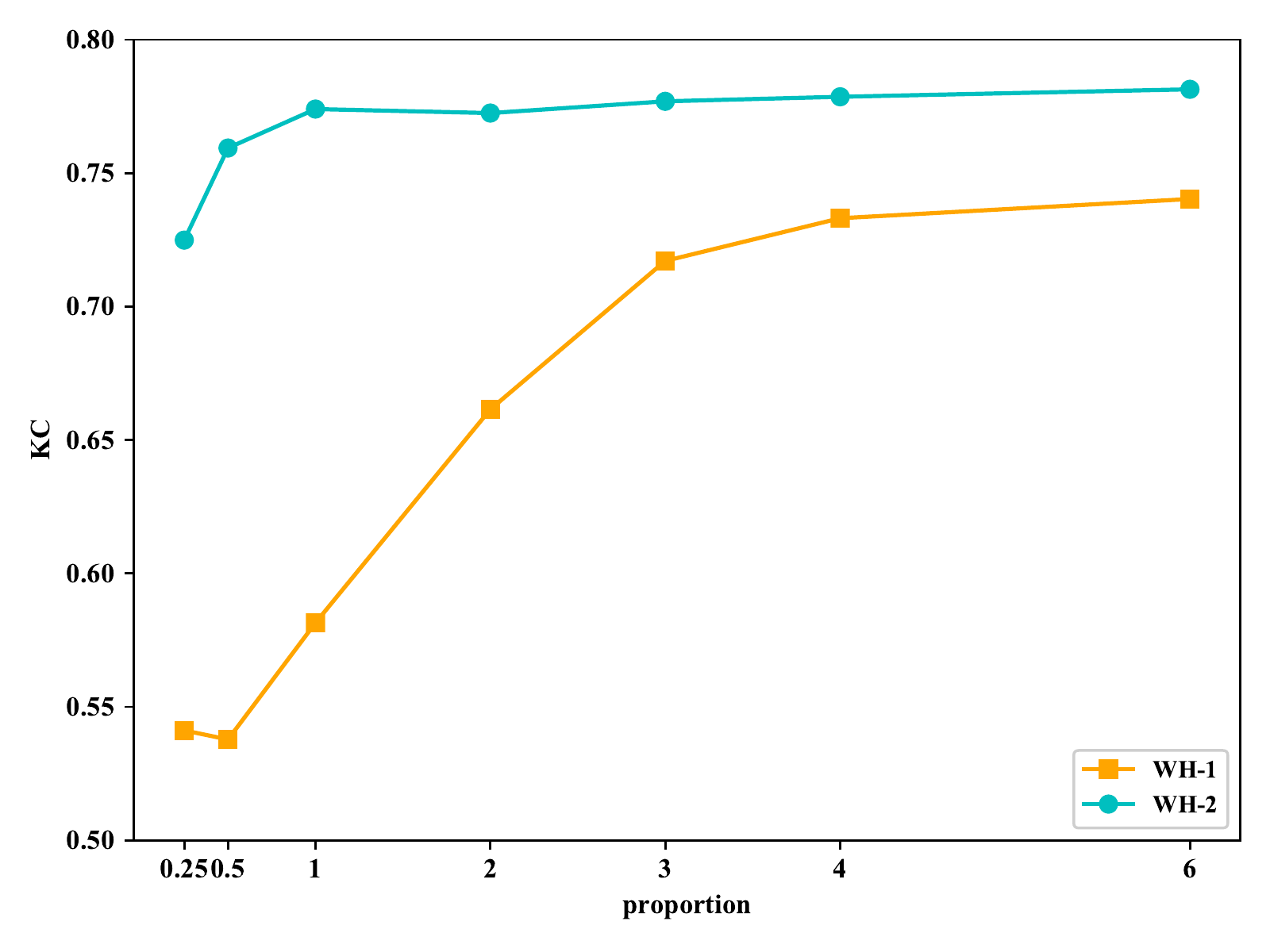}
  \caption{Relationship between the performance of the DSMS-CN and proportion of non-change and change classes in training samples.}
  \label{fig_propor}
\end{figure}
\par Besides, the proportion of non-change and change classes in training samples also affects the performance of the proposed method. The relationship between them over two data sets is shown in Fig. \ref{fig_propor}. Specifically, on the WH data set, the number of samples in change class generated by pre-classification is not many, so if the proportion of two classes is small, the number of non-change class in the training samples are too few training samples, causing an unsatisfactory change detection result. As the proportion increases, training samples contain enough unchanged land-cover types, hence the performance is improved. When the proportion is larger than 4, the performance becomes stable. Considering the computational expense, the proportion over WH data set is set to 4. The performance tendency on the HY data sets is similar. However, a sufficient number of samples in change class can be generated by pre-classification on the HY data set, hence the performance is acceptable even if the proportion is only 0.25. When the proportion is larger than 1, the performance of the DSMS-CN becomes stable. Thus on the HY data set, the proportion is set to 1.

\begin{figure*}[ht]
  \centering
  \subfloat[]{
    \includegraphics[width=2.2in]{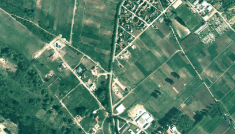}
  \label{Szada_1}}
  \hfil
  \subfloat[]{
    \includegraphics[width=2.2in]{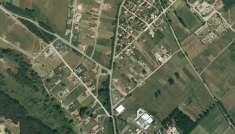}
  \label{Szada_2}}
  \hfil
  \subfloat[]{
    \includegraphics[width=2.2in]{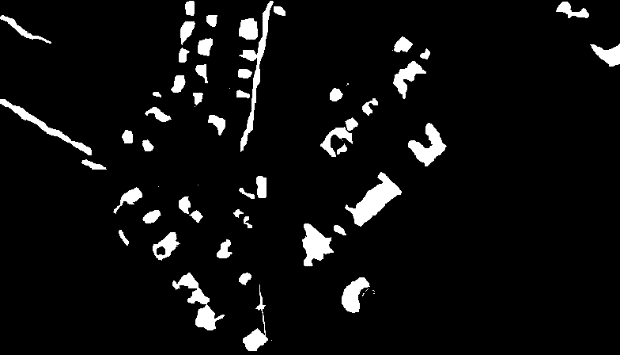}
  \label{Szada_GT}}

  \subfloat[]{
    \includegraphics[width=2.2in]{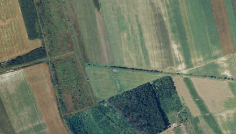}
  \label{Tiszadob_1}}
  \hfil
  \subfloat[]{
    \includegraphics[width=2.2in]{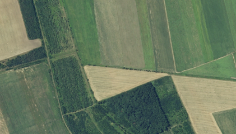}
  \label{Tiszadob_2}}
  \hfil
  \subfloat[]{
    \includegraphics[width=2.2in]{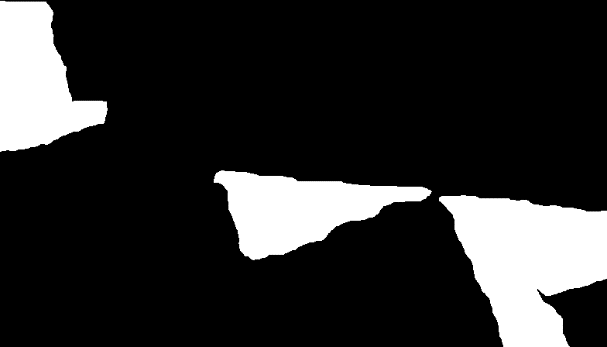}
  \label{Tiszadob_GT}}
  \hfil

  \caption{The multi-temporal images selected as the test set from the ACD data set. (a) Pre-change of Szada-1. (b) Post-change of Szada-1. (c) Ground truth. (d) Pre-change of Tiszadob-3. (e) Post-change of Tiszadob-3. (f) Ground truth.}
  \label{ACD_test_fig}
\end{figure*}

\section{Supervised Change Detection Experiment}\label{sec:5}

\subsection{ACD Data Set}
In order to train the proposed network and evaluate our method, the SZTAKI AirChange Benchmark set (ACD) is employed \cite{Benedek2009, Benedek2008}. The data set contains three sets of registered multi-temporal RGB aerial image pairs acquired in different seasonal conditions and associated ground truth. The size of each image is 952$\times$640 and their spatial resolution is 1.5m. The main differences between image pairs are new built-up regions, fresh plough-land and groundwork before building over. As a public data set, ACD has already been used in \cite{Zhan2017, CayeDaudt2018, Benedek2009, Benedek2008}.

\subsection{Experiment Settings}
\par First, to augment available training data, all possible flips and rotations of multiple of 90 degrees are used. For the data set split, we adopt the way proposed in \cite{Zhan2017} and \cite{CayeDaudt2018}: the top-left 784x448 corner of the Szada-1 and Tiszadob-3 are cropped for testing, and the rest of the images are used for training. The test image pairs is shown in Fig. \ref{ACD_test_fig}. Szada and Tiszadob are treated separately into two different data sets, and the images named “Archieve” are ignored, since it contains only one image pair. 

\par In the supervised algorithm, “he-normal” way is used to initialize the weights and bias of DSMS-FCN. To overcome the skew-class problem, the loss function of DSMS-FCN applies WBCE. Adam optimizer is chosen to train the network (learning rate is set to 2e-4). Dropout is used to avoid overfitting during the training phase. As proposed in \cite{Krahenbuhl2011a}, the weights and parameters of the Gaussian kernel of FC-CRF are determined by grid search on training set and the penalty of pairwise potential is trained by L-BGFS algorithm.

\par To evaluate the effectiveness of the proposed models, six state-of-the-art methods are used for comparison, including DSCN \cite{Zhan2017}, CXM \cite{Benedek2009}, SCCN \cite{Liu2018} and three fully convolutional networks proposed in \cite{CayeDaudt2018}. The three comparative fully-convolution networks are FC-EF, FC-Siam-Conc, FC-Siam-Diff. The experimental results of the comparison methods use the values obtained in the literature \cite{Zhan2017} and \cite{CayeDaudt2018}. For the purpose of better reflecting the role of FC-CRF, we separately evaluate the results obtained by DSMS-FCN and the combination of DSMS-FCN and FC-CRF. In addition, in order to verify the superiority of MFCU compared to conventional single-scale convolution unit, we modify the fully convolution architecture FC-EF proposed in \cite{CayeDaudt2018} and replace the 3$\times$3 convolution kernel with our MFCU. The modified network is denoted as MSFC-EF. Same as the unsupervised experiment, precision rate, recall rate, OA, F1 score, and KC are employed as evaluation criteria.

\begin{figure*}[!t]
  \centering
  \subfloat[]{
    \includegraphics[width=1.6in]{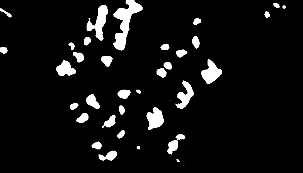}
  \label{Szada_MSFC-EF}}
  \hfil
  \subfloat[]{
    \includegraphics[width=1.6in]{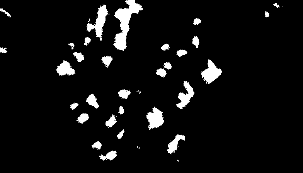}
  \label{Szada_MSFC-EF-FCCRF}}
  \hfil
  \subfloat[]{
    \includegraphics[width=1.6in]{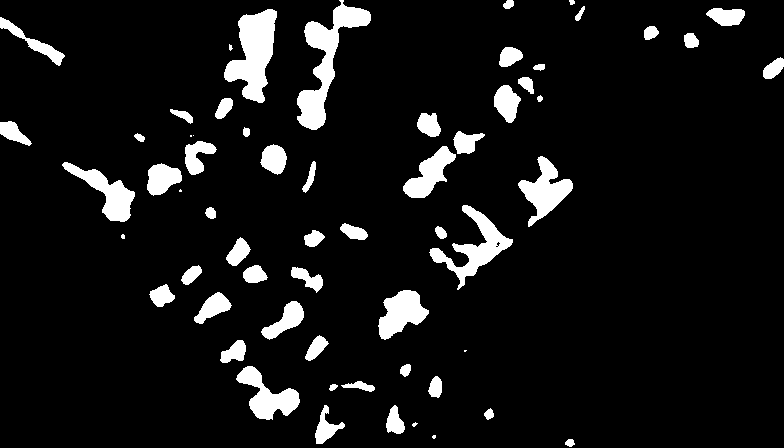}
  \label{Szada_DSMS-FCN}}
  \hfil
  \subfloat[]{
    \includegraphics[width=1.6in]{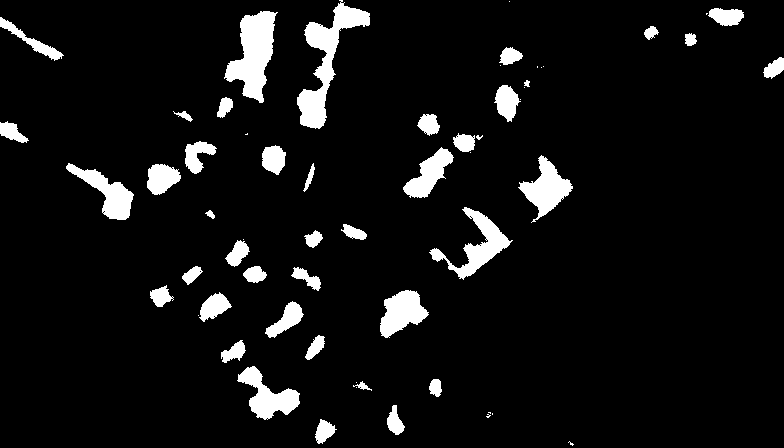}
  \label{Szada_DSMS-FCN-FCCRF}}
  \caption{Change detection results obtained by different methods on the Szada-1 of ACD. (a) MSFC-EF. (b) MSFC-EF-FCCRF. (c) DSMS-FCN. (d) DSMS-FCN-FCCRF.}
  \label{Szada_result}
\end{figure*}

\begin{figure*}[!t]
    \centering
    \subfloat[]{
      \includegraphics[width=1.6in]{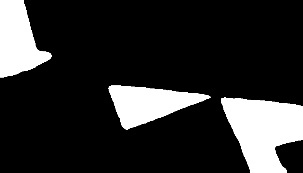}
    \label{Tiszadob_MSFC-EF-FCCRF}}
    \hfil
    \subfloat[]{
      \includegraphics[width=1.6in]{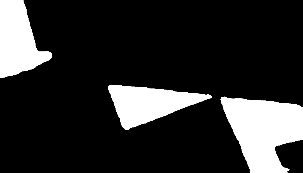}
    \label{Tiszadob_MSFC-EF}}
    \hfil
    \subfloat[]{
      \includegraphics[width=1.6in]{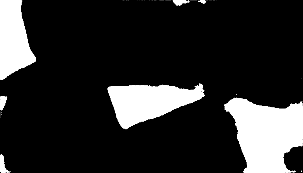}
    \label{Tiszadob_DSMS-FCN}}
    \hfil
    \subfloat[]{
      \includegraphics[width=1.6in]{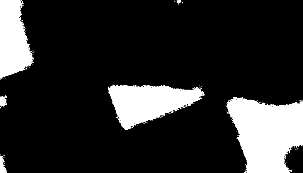}
    \label{Tiszadob_DSMS-FCN-FCCRF}
  }
    \caption{Change detection results obtained by different methods on the Tiszadob-3 of ACD. (a) is MSFC-EF. (b) is MSFC-EF-FCCRF. (c) is DSMS-FCN. (d) is DSMS-FCN-FCCRF.}
    \label{Tiszadob_result}
\end{figure*}

\subsection{Experimental Result and Analysis}
Fig. \ref{Szada_result} and Fig. \ref{Tiszadob_result} are illustrations of our results on the ACD data set. We could see that both our DSMS-FCN and improved FC-EF network, namely MSFC-EF, can achieve satisfactory qualitative results. Combined with FC-CRF, more low-level information is considered, thus the results obtained by deep fully convolutional network are refined.

Table \ref{Szada_table} and Table \ref{Tiszadob_table} report the quality analysis results on two test image pairs. The results obtained on the Szada-1 show the superiority of the proposed network, which obviously outperforms all the other comparison methods in recall metric, F1 score, and OA. Utilizing MFCU, each metric of MSFC-EF is better than FC-EF. Combined with FC-CRF, OA, F1 score and KC of MSFC-EF and DSMS-FCN are further improved. The DSMS-FCN-FCCRF achieves the best recall metric, OA, F1 score and KC.

\begin{table}[!t]
  \captionsetup{font={small}}
  \renewcommand{\arraystretch}{1.20}
  \caption{ACCURACY ASSESSMENT ON THE BINARY CHANGE MAPS OF DIFFERENT METHODS ON THE SZADA-1 OF ACD DATA SET}
  \label{Szada_table}
  \centering
  \begin{tabular}{c c c c c c}
  \hline
  \bfseries \textbf{Method} & \bfseries \textbf{Pre.} & \bfseries \textbf{Rec.} &  \bfseries  \textbf{OA}& \bfseries \textbf{F1} &\bfseries \textbf{KC}\\
  \hline\hline
  \textbf{DSCN}	& 0.412	& 0.574		& NA & 0.479	& NA \\ 	
  \textbf{CXM}	& 0.365	& 0.584		& NA & 0.449	& NA \\  									
  \textbf{SCCN}	& 0.224 & 0.347		& NA & 0.287	& NA \\ 									
  \hline
  \textbf{FC-EF} & 0.4357	& 0.6265		& 0.9308 & 0.5140	& NA \\  								
  \textbf{FC-Siam-Conc}	& 0.4093	& 0.6561		& 0.9246 & 0.5041	& NA \\ 	 							
  \textbf{FC-Siam-Diff}	& 0.4138	& \textbf{0.7238}		& 0.9240 & 0.5266	& NA \\  									
  \hline
  \textbf{MSFC-EF}	& 0.4725	& 0.6237	&  0.9374	&	0.5377 & 0.5048 \\ 		 							
  \textbf{MSFC-EF-FCCRF}	& 0.4888	& 0.6014 	& 0.9400 & 0.5393	& 0.5076 \\ 									
  \textbf{DSMS-FCN}	& 0.4835	& 0.6753	& 0.9392	& 0.5635	& 0.5317 \\
  \textbf{DSMS-FCN-FCCRF}	& \textbf{0.5278}	& 0.6339 & \textbf{0.9457} & \textbf{0.5772}	& \textbf{0.5484} \\
  \hline
  \end{tabular}
  \end{table}

\par On the Tisza-3, though the performance of the DSMS-FCN is not the best, it is still superior to DSCN, CXM, SCCN, and other two FCNs. FC-EF achieves a very high F1 score of 0.9340, but MSFC-EF utilizing MFCU still obtain improvement of performance. All metric of MSFC-EF is the better than FC-EF. Adopting FC-CRF, OA, F1 score, and KC of MSFC-EF and DSMS-FCN are further improved. The MSFC-EF-FCCRF achieves the best performance. 

\par Depending on FC-CRF, the results obtained by deep FCN can be further improved. It is worth noting that in the two test image-pairs, the performance of training  architectures is obviously better than the patch-based method DSCN and other architectures. 

\par
\begin{figure}[t]

  \centering
  \includegraphics[scale=0.65]{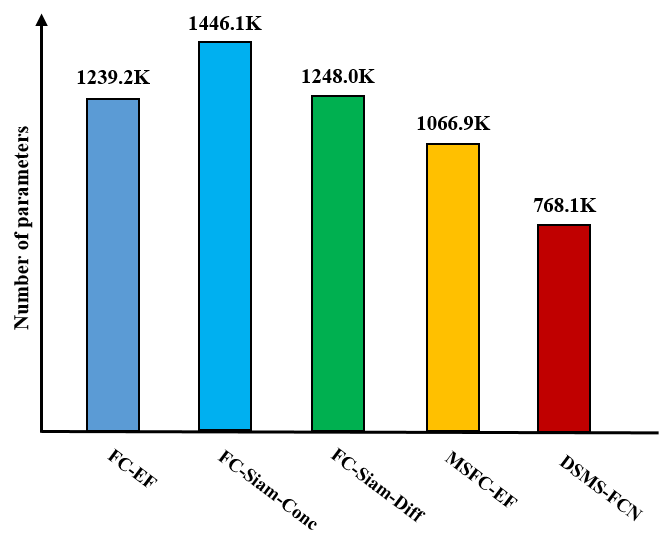}
  \caption{Comparison of the five FCN architectures in terms of model size.}
  \label{super_parameters}
\end{figure}

 The numbers of total trainable parameters in the five FCN architectures are shown in Fig. \ref{super_parameters}. The number of parameters in the proposed DSMS-FCN is in the smallest quantity. Compared to the FC-Siam-Conc and FC-Siam-Diff, the total number of parameters in DSMS-FCN is reduced by about 46.9$\%$ and 38.4$\%$, respectively. However, DSMS-FCN shows obviously better performance on both data sets, which implies that the proposed network has both better change detection ability and smaller computational cost. Besides, the model size of MSFC-EF is reduced by about 14.5$\%$ compared to the original network FC-EF. Meanwhile, the MSFC-EF can outperform the original model FC-EF in the data of Szada-1 and Tisza-3. In contrast to the conventional single-scale convolution unit, MFCU has fewer parameters and more powerful feature extraction ability, which can efficiently improve the performance of the deep network.

Furthermore, the inference time of DSMS-FCN is below 0.1s per image and the inference time of FC-CRF is under 1s per image, which means our method can efficiently process multi-temporal VHR images in real-time.

\section{Conclusion}\label{sec:6}

\par This paper utilizes a powerful multi-scale feature convolution unit for change detection in VHR images. Differing from conventional single-scale convolution unit that only extracts single-scale features in one layer, MFCU is capable of extracting multi-scale spatial-spectral features in the same layer by four ways. Specifically, the 1$\times$1 convolution kernel focus on extracting the features of a pixel itself. The 3$\times$3 convolution kernel can extract spatial-spectral features. The 5$\times$5 convolution kernel extracts larger-scale features. The 3$\times$3 max pooling is able to extract the most salient features and alleviate the smoothing effect of the convolution operation.

\par Based on MFCU, two novel deep siamese convolutional neural networks are designed for unsupervised and supervised change detection. In unsupervised change detection, DSMS-CN is trained on the samples generated by automatic pre-classification. In supervised change detection, DSMS-FCN is capable of processing images of any size and does not require sliding patch-window, thus the accuracy and inference speed could be significantly improved. To overcome the inaccurate localization problem, the FC-CRF is adopted to refine the results obtained by DSMS-FCN. Through using the output of DSMS-FCN as unary potential, the FC-CRF is combined with DSMS-FCN. 

\par On the unsupervised change detection experiments with the two challenging data sets, the experimental results indicate that DSMS-CN outperforms the other comparison methods with better F1 score, OA, and KC, which confirms the powerful extraction ability of MFCU and outstanding fitting capability of DSMS-CN. In the supervised change detection experiments with ACD data set, compared with three FCNs and other state-of-the-art methods, DSMS-FCN delivers a competitive performance. And MFCU can be treated as a general module and embedded into deep CNN by replacing the original single-scale convolution unit to improve the performance. What's more, FC-CRF does make the results obtained by deep FCNs more accurate. Lastly, the inference time of our supervised architecture is below 1s per image, thus it can predict change maps of multi-temporal VHR images in real-time.

\par Our future work will focus on applying deep siamese architecture for change detection in heterogeneous VHR images and utilizing FC-CRF to improve the performance of traditional change detection methods in VHR images.

\begin{table}[t]
  \captionsetup{font={small}}
  \renewcommand{\arraystretch}{1.20}
  \caption{ACCURACY ASSESSMENT ON THE BINARY CHANGE MAPS OF DIFFERENT METHODS ON THE Tiszadob-3 OF ACD DATA SET}
  \label{Tiszadob_table}
  \centering
  \begin{tabular}{cccccc}
  \hline
  \bfseries \textbf{Method} & \bfseries \textbf{Pre.}  & \bfseries \textbf{Rec.}  & \bfseries \textbf{OA} & \bfseries \textbf{F1} & \bfseries \textbf{KC}\\
  \hline\hline
  \textbf{DSCN}	& 0.883	& 0.851	&  NA	& 0.867	& NA \\ 	
  \textbf{CXM}	& 0.617	& 0.934	&  NA	& 0.743	& NA \\  									
  \textbf{SCCN}	& 0.927 & 0.798	&  NA	& 0.858	& NA \\ 									
  \hline
  \textbf{FC-EF} & 0.9028	& 0.9674	&  0.9766	& 0.9340	&NA \\  								
  \textbf{FC-Siam-Conc}	& 0.7207	& 0.9687	&  0.9304	& 0.8265	& NA \\ 	 							
  \textbf{FC-Siam-Diff}	& 0.6561	& 0.8829	&  0.9137	& 0.7778	& NA \\  									
  \hline
  \textbf{MSFC-EF}	& 0.9489	& 0.9763		& 	0.9870 & 0.9624 & 0.9545 \\ 		 							
  \textbf{MSFC-EF-FCCRF}	& \textbf{0.9514}	& \textbf{0.9765} 	& \textbf{0.9874} & \textbf{0.9638}	& \textbf{0.9560} \\ 									
  \textbf{DSMS-FCN}	& 0.8691	& 0.8690	&  0.9552	& 0.8690	& 0.8420 \\
  \textbf{DSMS-FCN-FCCRF}	& 0.8918	& 0.8856 & 0.9620	& 0.8886	& 0.8658 \\
  \hline
  \end{tabular}
\end{table}

%\section*{Acknowledgment}

%This work was supported in part by the National Natural Science Foundation of China under Grant 61971317, 41801285, 61822113 and 41801285.

\ifCLASSOPTIONcaptionsoff
  \newpage
\fi

\bibliographystyle{IEEEtran}
\bibliography{DSMSCN.bib}

\end{document}